\newif \ifformat
\formatfalse 

\ifformat
\documentclass{statsoc}
\else
\documentclass[a4paper, 12pt]{article}
\fi

\usepackage[top=1.7cm, left=1.2cm, right=1.3cm, bottom=2cm]{geometry}

\ifformat
\else
\usepackage[pdfauthor={Tuomas A. Rajala}, hidelinks, colorlinks, citecolor=blue]{hyperref}
\fi

\usepackage{graphicx}
\usepackage[utf8x]{inputenc}
\usepackage{natbib}
\usepackage{amsmath,amsfonts}

\usepackage{color}

\newcommand{\rv}{\mathbf{r}}
\newcommand{\x}{\mathbf{x}}
\newcommand{\s}{\mathbf{s}}
\newcommand{\g}{\mathbf{g}}
\newcommand{\vv}{\mathbf{v}}

\newcommand{\z}{\mathbf{z}}

\newcommand{\ov}{\mathbf{o}}

\newcommand{\R}{\mathbb{R}}
\newcommand{\E}{\mathbb{E}}

\newcommand{\T}{T}

\ifformat
\title[Multivariate Gibbs models and variable selection]{Detecting multivariate interactions in spatial point patterns with Gibbs models and variable selection}
\author{T Rajala\footnote{\textit{Corresponding author. Postal address: University College London, Gower Street, London WC1E 6BT, United Kingdom. E-mail: \texttt{t.rajala@ucl.ac.uk}}
}}
\address{University College London, London, UK}
\author{D J Murrell}
\address{University College London, London, UK}
\author[Rajala, Murrell, Olhede]{S Olhede}
\address{University College London, London, UK}

\else
\title{Detecting multivariate interactions in spatial point patterns with Gibbs models and variable selection.}
\author{T Rajala, D J Murrell, S Olhede\\University College London, London, UK}

\fi

\begin{document}

\maketitle

\begin{abstract}
We propose a method for detecting significant interactions in very large multivariate spatial point patterns. This methodology thus develops high dimensional data understanding in the point process setting.
The method is based on modelling the patterns using a flexible Gibbs point process model to directly characterise point-to-point interactions at different spatial scales. By using the Gibbs framework significant interactions can also be captured at small scales. 
Subsequently, the Gibbs point process is fitted using a pseudo-likelihood approximation, and we select significant interactions automatically using the group lasso penalty with this likelihood approximation. Thus we estimate the multivariate interactions stably even in this setting.
We demonstrate the feasibility of the method with a simulation study and show its power by applying  it to a large and complex rainforest plant population data set
of 83 species.
\end{abstract} 

\section{Introduction}
 
Spatial point patterns are a common form of observation in plant ecology~\citep{Waagepetersen2015b}, epidemiology~\citep{diggle2005nonparametric}, astrophysics~\citep{stoica2007three}, seismology~\citep{schoenberg2003multidimensional}, social science~\citep{amburgey1986multivariate}, medicine~\citep{Olsbo2013} and criminology~\citep{mohler2011self}. 
While understanding single, univariate spatial point patterns and their generating point processes is important, frequently we observe {\em labelled} point processes, or more precisely, multiple types of points. The prevalence of multivariate point processes is particularly noticeable in plant ecology where there may be many tens or hundreds of types (species) ~\citep{Flugge2014,baldeck2013soil,baldeck2013habitat,Kanagaraj2011,punchi2013effects}. Such processes have seen much less study in the statistical literature than univariate processes, and present a number of novel challenges, as we shall explain and address in this paper.  

To be able to make sense of multivariate point processes, 
we focus on addressing three important outstanding problems in understanding interactions: i) characterizing patterns associated with both small and large scales simultaneously, ii) characterizing multiple features spanning more than one variable, and iii) estimating such patterns stably. The effect of including these characteristics in our studies of a multivariate point process greatly increases the number of potential parameters to characterise and naturally leads to difficulties in arriving at unique and stable parameter estimates.  

Early approaches to modelling and analysing more than one point processes focused on bivariate representation and analysis~\citep{diggle1983bivariate,amburgey1986multivariate,Brix2001,gelfand2004nonstationary,shimatani2001multivariate,diggle2005nonparametric}, and going much  beyond the bivariate case has proved very challenging. To the best of our knowledge the most diverse model-based analysis of multivariate point patterns to date is of nine rainforest tree species by~\cite{Waagepetersen2015b}. These authors used a multivariate log-Gaussian Cox process model, and fitted the model using non-parametric least squares.  
The reason given for confining studies to only nine species by the previous authors was to limit the computational burden of analysis. This work is inspirational, but our motivating problem in this paper is to jointly analyse an order of magnitude more species. More specifically, there are 300 species (types) in the full dataset used in ~\cite{Waagepetersen2015b}, and we wish to extend analysis to investigate as many of these species as possible. This brings us into the realm  of high dimensional statistics as the number of interactions scales exponentially in the number of species, whilst the number of points only scales linearly in the number of species. To deal with this inconvenient scaling we shall need to use shrinkage, as is commonly done in high dimensional data analysis, and has already been developed for regression problems and covariance estimation, see e.g. ~\cite{van2011statistics}.
In the context of  point processes, because estimation is not implemented with linear methods, penalization needs to be deployed carefully. Starting from ideas of~\cite{Baddeley2014} we shall use a generalized regression--based fitting approach, and so we borrow ideas from second generation penalized regression rather than from matrix shrinkage, even if we are estimating co-associations rather than a mean intensity. 

A second significant problem in point process modelling is proposing models, and associated estimation methods, that yield sufficient multiscale behaviour, e.g. variability at fine scales, as well as over medium to long scales. It is all very well to posit variability for fine scales using a log-Gaussian Cox process, but as estimation is normally based on some form of averaging, unless the random intensity is very high locally, it will be impossible to estimate the log-Gaussian Cox process' generating mechanism as we will not have enough points.
For longer spatial scales the log-Gaussian Cox process is a well-suited modelling framework, but it is not a good framework for studying small-scale interactions. Instead we shall use the multivariate Gibbs point process model to discover small scale point-to-point interactions in the same Barro Colorado Island (BCI) rainforest dataset studied by~\cite{Waagepetersen2015b}. 

Key to our modelling and estimation is therefore capturing an appropriate degree of sample heterogeneity.
The estimation framework we shall introduce is able to take into account variations associated with i) habitat associations, i.e. correlation of species presence in the landscape with known environmental covariates, with ii) dispersal mechanisms and competition such as seedling clustering or self-thinning, and with iii) attraction and repulsion between the small scale locations between different species. To demonstrate utility, we will fit the model to an adult plant community consisting of $83$ species, an order of magnitude more species than~\cite{Waagepetersen2015b}. 

The analysis of multivariate point patterns of more than a handful of species \citep[usually two, e.g.][]{Brix2001,diggle1983bivariate,Hogmander1999,Funwi-Gabga2012} has mostly relied on non-parametric estimation techniques. Many ecological analyses of large and diverse rainforest datasets have been carried out using the $K$-function or similar non-parametric summaries, either directly by comparing the summary values under various null-model scenarios, or indirectly as part of minimum contrast model fitting of Cox processes \citep[e.g.][]{Lan2012, Flugge2014, Waagepetersen2015b, Yang2016a,Velazquez2016, Brown2016}. In the Gibbs-model framework, the three-variate analysis of different size trees by \cite{Grabarnik2009} is perhaps the most extensive modelling approach and closest in spirit to the work of this paper. We will extend their trivariate case to a full multi-scale multivariate interaction Gibbs model.

Direct likelihood inference for Gibbs models is not available due to intractable normalising constants, but several pseudo-likelihood approximations are available, such as the ``Berman-Turner machine'' \citep{Baddeley2000a} used by \cite{Grabarnik2009} which casts the pseudo-likelihood estimation equation as a Poisson regression problem, and subsequently estimates the model with standard statistical software. The Poisson regression approach is surpassed in accuracy by the logistic regression approach developed by \cite{Baddeley2014}, who formulates the pseudo-likelihood estimation equation as a logistic regression using auxiliary dummy point configurations. Again, very conveniently, standard statistical software can be used to fit the model.

With a model and the likelihood approximation at hand, we will next tackle the issue of high dimensional variable selection: Any reasonable model for a $p$-variate point pattern will have a high number of parameters when $p\gg 3$, scaling at least like ${\cal O}(p^2)$. As an illustration, the model we present estimates intra-species interactions and pairwise species-to-species interactions, on 3 different spatial scales, and includes 6 covariates, giving a total 
of $\approx 11,000$ parameters, to be compared with the number of observations of $31,650$ points, making the number of parameters and observations on the same order. To discover significant interactions in such a high dimensional setting, we will use recent research on penalized optimization. Several techniques are available, and to show proof of concept we will be using the group lasso \citep{yuan2006model,Meier2008}. We will also perform a limited comparison to the Bayesian spike and slab variable selection approach \citep{Mitchell1988}. 
Using penalized regression (the group lasso) rather than 
full posterior inference to compute a MAP yields both simplicity in interpretation and computational speed. This allows us to study several ``priors'', or penalization choices, at once and thus this allows us to make fewer assumptions on the generating mechanism of the data, in fact letting us explore the properties of our modelling framework.

We start in Section 2 by introducing the Gibbs model, then recall the chosen likelihood approximation and discuss the chosen variable selection techniques in detail. In Section 3 we recall for comparison the non-parametric Monte Carlo technique often used for analysing $p$-variate patterns when $p\gg 1$. In Section 4 we test the method on several increasingly complex simulation scenarios, to get a better understanding of the performance of the method. In Section 5 we apply the method to the Barro Colorado Island rainforest data~\citep{Condit1998}, that inspired these developments. We conclude in Section 6, and discuss outstanding problems and future avenues of investigation.

\section{The Gibbs model and Gibbs model fitting with variable selection}

Let the observed multivariate point pattern be a set of labelled point locations $\x = \{(x,t)\}$ where $x\in W$ are the observed locations inside a known, bounded observation window $W\subset\R^d$, $d=1,2,3,...$ (for the problems we shall study, $d=2$), 
and $t\in \{1,...,p\}$, $p\ge 1$, are categorical labels (types, or in our case, species) attached to each location. Denote the type $i$ sub-pattern by $\x_i = \{x: (x,i)\in \x\}$. Write $n_i(A)=\#(\x_i\cap A)$ for the count of points of type $i$ in a set $A\subset\R^d$, and write $n_i=n_i(W)$. As is standard let $b(u,r)$ denote a ball of radius $r$ centred at $u\in \R^d$.

\subsection{Gibbs models for point patterns}
Our model for multivariate spatial point patterns is part of the Gibbs point process model family; we refer to \cite{vanLieshout2000} who details general properties of Markov Point Processes (where Gibbs point processes are a special case), 
and \cite{Illian2008a} and \cite{Chiu2013} who discuss general point processes. We will use a Gibbs model, defined from a set of {\em potential functions} $\phi_{ij}$, to have a probability density of the form
\begin{equation}
f(\x) \propto \exp[\sum_{i,j=1}^p\phi_{ij}(\x_i,\x_j)],
\label{eq:gibbs}
\end{equation}
with respect to the unit rate Poisson process $\mu_1$ on $W$. The normalising constant for the density,
\[
\int_{\mathcal{X}}\exp[\sum_{i,j=1}^p\phi_{ij}(\x_i,\x_j)]\mu_1(d\x),
\]
with $\mathcal{X}$ the space of all locally finite multitype point patterns, is in practice intractable for all but the homogeneous Poisson models where $\phi_{ij}\equiv \alpha_{i}\in\R$. The functions $\phi_{ij}$ are used to specify the model class member. We will use the special form of
\begin{eqnarray}
\phi_{ij}(\x_i,\x_j) & = & \left\{\begin{array}{lc}
\sum_{x\in \x_i} \alpha_i^\T \z_i(x) +\ \sum_{x\in \x_i} \beta_{ii}^\T \g_{ii}(x, \x_i\setminus x)&, i=j\\[1em]
\sum_{x\in \x_i} \beta_{ij}^\T \g_{ij}(x, \x_j)&, i\neq j
\end{array} 
\right. .
\label{eq:gibbs2}
\end{eqnarray}

In~\eqref{eq:gibbs2} the parameters $\alpha_i\in \R^{K_i+1}$ regulate intensity and covariate effects (so-called first order effects), and $\beta_{ij}\in \R^{K_{ij}}$ are parameters for interactions between the locations (second order effects). The $\beta_{ij}$'s give the magnitudes of the interactions, whereas the vector-valued functions $\g_{ij}$, which we specify later in Section \ref{sec:interaction}, determine the form, spatial scales and orders of the interactions. The covariate effects $\z_i(u) = [1\ z_{i1}(u)\ z_{i2}(u) \ldots z_{iK_i}(u)]^\T$ represent the baseline effect and any covariate and trend effect values we have at locations $u\in W$. In this formulation we assume that the covariates are available everywhere in the window $W$. This is usually achieved by interpolation from prior data collection efforts. 

To get a heuristic understanding of the model, assume first that all parameters except $\alpha_i\in \R$ are zero and $\z_i\equiv 1$. Then the density \eqref{eq:gibbs} becomes $\exp[\sum_{i=1}^{p}n_i\alpha_{i}]=\prod_{i=1}^{p}e^{n_i\alpha_{i}}$, which is the likelihood of a collection of $p$ independent homogeneous Poisson processes. Now add some non-constant covariates $z_{i2}(u), z_{i3}(u),\ldots$ and set $\alpha_{i2}, \alpha_{i3}, \ldots \neq 0$: The model becomes a collection of independent and inhomogeneous Poisson processes. We subsequently add intra-type interaction terms by letting $\beta_{ii}\neq 0$: The independent components are no longer Poisson, but exhibit internal, within type point-to-point interactions (attraction or repulsion depending on the sign of $\beta_{ii}$). Finally we can add inter-type interaction terms by setting $\beta_{ij}\neq 0$: The locations of different types are no longer independent.

The model of~\eqref{eq:gibbs} is log-linear in parameters $\alpha$ and $\beta$. We collect the covariate parameters in the vector $\theta_0 = [\alpha_1^\T \ldots \alpha_p^\T]^\T$, the intra-type interaction parameters to $\theta_1=[\beta_{11}^\T\ldots\beta_{pp}^\T]^\T$ and inter-type interaction parameters, assuming for now that $i$ and $j$ interact symmetrically, to $\theta_{2}=[\beta_{12}^\T\ldots \beta_{(p-1)p}^\T]^\T$. Subsequently we collect them in the vector $\theta = [\theta_0^\T \ \theta_1^\T \ \theta_2^\T]^\T$. Then the density in (\ref{eq:gibbs}) can be written in the form
\begin{equation}
f(\x) = f_\theta(\x) \propto\exp[\theta^\T\mathbf{v}],
\label{eq:log-linear}
\end{equation}
where the vector $\mathbf{v}=\mathbf{v}(\x)=[\s_0^\T\ \s_1^\T\ \s_2^\T]^\T$ has components
\begin{itemize}
\item $\s_0 = [\s_1^\T\ldots \s_p^\T]^\T$ with $\s_i=\sum_{x\in \x_i}\z_i(x)$
\item $\s_1=[\s_{11}^\T\ldots\s_{pp}^\T]^\T$ with $\s_{ii} = \sum_{x\in \x_i}\g_{ii}(x,\x_i\setminus x)$
\item $\s_2=[\s_{12}^\T\ldots\s_{(p-1)p}^\T]^\T$ with $\s_{ij} = \sum_{x\in \x_i}\g_{ij}(x,\x_j)$, $i<j$.
\end{itemize}
The vector $\mathbf{v}=\mathbf{v}(\x)$ is written as a matrix $[\vv^\T]$, and takes the role of a design matrix in the standard regression setting. The matrix has only one row when the point process is observed only once. Independent replicates would result in additional rows in the matrix.

Given data $\x$, covariates $\z$ and the interaction functions $\g_{ij}$, the vector $\vv$ is fixed and the non-normalised model (\ref{eq:log-linear}) is log-linear in the unknown coefficients $\theta$. The model therefore belongs to the family of exponential Gibbs models, and we can apply inference techniques designed for the exponential Gibbs family. 

\subsection{The interaction functions}
\label{sec:interaction}
\newcommand{\inth}{\psi}

We now define the exact model we will use in our examples. Several definitions are available for the interaction functions $\g_{ij}$ in (\ref{eq:gibbs2}). We assume $\g_{ij}$ are non-negative functions to remove sign ambiguity when estimating $\beta_{ij}$. The most popular class of models is the pairwise interacting models with
\[
\g_{ij}(x,\x\setminus x) = \sum_{y\in \x\setminus x} \inth_{ij}(||x-y||),
\]
for some functions $\inth_{ij}:\R_+\mapsto\R_+^{K_{ij}}$. The most common choice in the case of $K_{ij}=1$, is the Strauss model 
\begin{eqnarray}
	g(x,\x\setminus x) &=& \sum_{y\in \x\setminus x} 1(||x-y|| < r),\;r>0,
\end{eqnarray}
effectively counting the number of $r$-close pairs in the pattern. In the univariate case the Strauss model is only valid when the corresponding interaction coefficient $\beta<0$ so that fewer point pairs leads to a higher likelihood. Trying to model positive interactions, or clustering, leads to an unstable model that produces patterns of singular mega-clusters, so the case $\beta>0$ is excluded for the simple Strauss model \citep{Gates1986}.

To circumvent this limitation \cite{Geyer1999} introduced a model he called the saturation model that still defines a locally stable process even with a positive interaction parameter $\beta>0$.  
In the simple univariate case the saturation model is defined via 
\begin{eqnarray}
	g(x,\x\setminus x) &=& \min\left\{c, \#[(\x\setminus x)\cap b(x,r)]\ \right\},\;r>0,\;c
    \in{\mathbb{N}}_+,
\end{eqnarray}
where the range $r$ is the reach of the Euclidean neighbourhood, and $c$ is a saturation level. In this model, each point contributes to the likelihood a factor relative to the number of $r$-neighbours or $c$, which ever is smaller (hence the saturation). In ecological terms, the Geyer model is able to capture the fact that individuals may cluster at some distances, but are likely to segregate at shorter distances due to intense competition, and the saturation parameter reproduces the feature that the neighbourhood must eventually saturate with individuals as resources are finite. The model belongs to a class of interacting neighbour models \citep{Grabarnika2001}, so-named because the conditional intensity (\ref{eq:papangelou}) for this model depends not just on the local neighbourhood of a point in $u$ (which it does for pairwise models), but also on the neighbourhoods of the neighbours of $u$.

To model several types of points and more than one spatial scale, we generalise the models in two ways by adding (i) multiple ranges, and (ii) cross-type interactions when $p\ge 2$. 
Let $\rv_{ij}=\{r_{ijk}:0=r_{ij0}<r_{ij1}<...<r_{ijK_{ij}}\}$ be a fixed increasing vector of ranges for $i\le j$. Let $c_{ij}=\{c_{ijk}\in\mathbb{N}\}$ be the saturation parameters. For a given range vector, write 
\[
ne_{ijk}(x,\x):= \#\{\x \cap \ [b(x,r_{ijk})\setminus b(x,r_{ij(k-1)})] \},
\]
for the number of neighbours of $\x$ a point $x$ has in the annulus between ranges $r_{ij(k-1)}$ and $r_{ijk}$. Then the multi-step multi-type Strauss model is defined using the interaction function made of components
\begin{eqnarray}
	g_{ijk}(x, \x) &:=& ne_{ijk}(x,\x),
	\label{eq:Strauss}
\end{eqnarray}
and the multi-step multi-type saturation model of components
\begin{eqnarray}
	g_{ijk}(x, \x) &:=& \min[c_{ijk}, ne_{ijk}(x,\x)]
	\label{eq:saturation}
\end{eqnarray}
in definitions of $\g_{ij}$. Note that the Strauss model is a special case of the  saturation model with $c \rightarrow \infty$. 

Several other forms of $g_{ijk}$ can be used, and multiple forms can be combined as described in \cite{Baddeley2013c}. We will not pursue them here as either a) they can be approximated by the Strauss or saturation model as the interaction functions $\beta_{ij}^\T\g_{ij}$ are step-functions over spatial scales; b) there is not sufficient data available to estimate very fine details over many spatial scales; or c) they are computationally costly (e.g. morphological functions). 
However, if forms such as the area interaction model \citep{Baddeley1995a} seem more appropriate for a specific application, the proposed framework is still valid and can be adapted to be used in this setting. 

For the purpose of fitting the model (Section \ref{seq:2.2}), we need to define the conditional (or Papangelou) intensity of the model: At any point $u=(x,i) \in \R^d\times \{1,...,p\}$ let
\begin{eqnarray}
	\lambda_\theta( u ;\x) &:=& \frac{f_\theta(\x \cup u)}{f_\theta(\x \setminus u)} 
	\label{eq:papangelou}
\end{eqnarray}
with $0/0:=0$. Heuristically, $\lambda_\theta(u;\x)du$ can be understood as the conditional probability of observing a point $u$, given the rest of the pattern $\x$. For the exponential family Gibbs models the conditional intensity has the rather simple form of
\begin{eqnarray}
	\lambda_\theta( u ;\x) &=& \exp\left\{\theta^\T[\vv(x\cup u) - \vv(\x\setminus u)] \right\} = \exp[\theta^\T \vv(u;\x)],
\end{eqnarray}
where we use the notation $\vv(u;\x) = \vv(\x\cup u)-\vv(\x\setminus u)$. Notice that the intractable normalising constant cancels out in this expression.

The conditional intensity (\ref{eq:papangelou}) with the step-wise components (\ref{eq:Strauss} and \ref{eq:saturation}) is at any marked point $u=(x,i)$ 
\begin{eqnarray}
	\log \lambda_\theta(u;\x) &=& z(u)^\T\alpha_{i} + \sum_{i=1}^p\sum_{j=i}^p\sum_{k=1}^{K_{ij}}\beta_{ijk} \omega_{ijk}(u,\x_j), 
	\label{eq:papangelou2}
\end{eqnarray}
with 
\[
\omega_{ijk}((x,i),\x_j) =  g_{ijk}(x, \x_i\setminus x) + \sum_{y\in \x_j}[g_{ijk}(y, \x_i \cup x) - g_{ijk}(y, \x_i \setminus x)] .
\] 
The functions $\omega_{ijk}\in [-c_{ijk},n_ic_{ijk}]$ track the changes in the neighbourhood inclusion counts within and between different types. An illustration of the model is given in Appendix \ref{appendix:interaction}.

For both models the non-canonical parameters $r_{ijk}$ and $c_{ijk}$ cannot be directly inferred by the pseudo-likelihood methods. Thus the parameters need to be fixed as part of the model definition. In data analysis one usually has a priori information on relevant ranges $\{r_{ijk}\}$ (e.g. \citealt{Uriarte2004}), 
but the saturation level $\{c_{ijk}\}$ is harder to set. 
To reduce this complexity we propose to choose the saturation levels $\{c_{ijk}\}$ automatically depending on the abundances $n_j$. The idea is that if the abundance $n_j$ is high the neighbourhoods $ne_{ijk}$ and therefore $\omega_{ijk}$ saturate often even under the independence assumption. Under full independent assumption, the expectation of $\omega_{ijk}$ is a function of $c=c_{ijk}$ 
\[
t(c) := c[1-F_a(c-1)] + a[F_a(c-1) + F_a(c-2)],
\]
where $a = a_{ijk} = |b(o,r_{ijk})\setminus b(o,r_{ij(k-1)})|n_j/|W|$ and $F_a$ is the cumulative distribution function of a $Poisson(a)$ random variable (cf. Appendix \ref{appendix:saturation}). 
The function $t$ is non-decreasing and $t\rightarrow 2a$ when $c\rightarrow \infty$, so that $1-t(c)/2a$ is the cdf of the event that saturation occurs under independence. In order to avoid saturation due to high abundances alone, a sensible choice of $c$ is a value for which saturation under independence is unlikely. Therefore, we set a small $0< \epsilon < 1$ such that
\[
1 - t(c)/2a < \epsilon
\]
and, by further using the approximation $t(c)/2a \approx F_a(c-1)$,
we choose $c=c_{ijk}$ to be the $(1-\epsilon)$-quantile of $F_a$.
In the examples we use $\epsilon=.01$. Note that with the modification the interaction functions $\g_{ij}$'s become asymmetric, but we will treat the $\beta_{ij}$'s symmetrically in our examples for simplicity.

\subsection{Inference: Approximating the likelihood at its mode}
\label{seq:2.2}
The likelihood in Equation (\ref{eq:gibbs} equiv. \ref{eq:log-linear}) is not computationally tractable due to an unspecified normalising constant. In order to carry out standard likelihood inference the constant can be approximated by Monte Carlo techniques, but these tend to be computationally costly even for the univariate case. The more commonly used approach is to use pseudo-likelihood techniques, which replace the function to maximise with something that approximates the likelihood at its mode. We will use the recent developments proposed by \cite{Baddeley2014}, which in practice conveniently leads to a logistic regression formulation.


We have summarised the details of the method in Appendix \ref{appendix:likelihood}. For the purposes of the discussion, it suffices to know that the pseudo-likelihood function $\tilde{f}_\theta$ is formally a likelihood of a logistic regression function. It involves additional sets of random dummy points per type, of which construction, particularly their intensity $\rho_i$, is an additional user decision. \cite{Baddeley2014} discuss several potential options to be used for the dummy distributions and intensities. We will use the recommended homogeneous stratified uniform distributions and, if not otherwise stated, intensities that are four times the intensity of data.


To address boundary effects due to censoring near the edges of observation window $W$, we will exclude components of $\tilde{f}_\theta$, say $\mathbf{b}(u)$, for which the distance from $u$ to the border of $W$ is less than a range $r_{bor}>0$, i.e. in the sum in (Appendix \ref{appendix:likelihood} eq. \ref{eq:approxlikelihood}) the $W$ is replaced by $W\ominus b(o,r_{bor})$. The range $r_{bor}$ is taken to be the maximal interaction range in the model, and is determined by the $\g_{ij}$'s.

\subsection{Penalised inference for grouped coefficients}
\label{sec:penalised}

Our main goal is the detection and estimation of significant within-type and between-type interactions in multivariate point pattern data. To this end the previous section allows us to estimate groups of coefficients, such as the inter-type interaction vectors $\beta_{ij}\in \R^{K_{ij}}$ for type pairs $i\neq j$. Subsequent to that, we need to determine if $\beta_{ij}= 0$ or if this is not the case. 

In a general setting, let the length of an unknown coefficient vector $\theta$ be $M$, and split $\theta$ into $m$ smaller groups using a partition given by $\{\pi_1,...,\pi_m\}$ of $\{1,...,M\}$. For shorthand we will write $(g)=\pi_g$, where $g=1,...,m$. Write $e_g=0$ when $\theta_{(g)}=0$ and $e_g=1$ when at least one of the coefficients $\theta_{(g)}$ is non-zero. In our case the partition is given by the types and type-pairs, and the $e_g$'s are connected to the events $\beta_{ij}\neq 0$. Our task is to determine which $e_g$'s are non-zero in a given data $\x$, a task known as grouped variable selection~\citep{yuan2006model}. 


Variable selection and shrinkage in high dimension, usually implemented using penalised optimisation, has become an important computational statistical technique due to an increase in high data-throughput applications such as genomics and text analytics. 
Variable selection has already been applied to point pattern analysis, see for example  the work by \cite{Yue2014a}. The latter authors applied variable selection in the univariate case for covariate selection. Several methods exist for the particular problem of group-level selection (or penalisation) that we are addressing in this context~\citep[see e.g.][]{Breheny2009}. From the available selection of optimization criteria, we picked the group lasso for logistic regression \citep{Meier2008}, as a suitable penalized version of~\cite{Baddeley2014}. The group lasso is an extension of the original $\ell_1$-penalisation by \cite{Tibshirani1996} for individual coefficients, and has been further extended to a mixed level penalisation by \cite{Simon2013}. 
To be concrete, the group lasso estimator is defined by
\[
\hat\theta_\gamma = \text{argmax}_{\theta}\  L_\gamma(\x; \theta),
\]
with the group-penalised likelihood function
\begin{equation}
\label{eq:glasso}
L_\gamma(\x; \theta) := \log\tilde{f}_\theta(\x) - \gamma \sum_{g=1}^m|(g)|^{-\frac{1}{2}}\|\theta_{(g)}\|_2,
\end{equation}
where $\gamma>0$ is a penalisation parameter and $\|\cdot\|_2$ is the Euclidean distance. Some groups $\theta_{(g)}$ are shrunk to exactly 0, depending on the strength of the penalisation, so the group lasso does provide us with binary indicators $e_g = 1(\hat{\theta}_{(g)}\neq 0)$.

The penalisation parameter $\gamma$ is directly linked to the amount of non-zero $e_g$'s that the algorithm outputs and it needs to be chosen by the user. With some extra computational cost, we compute what is known as the lasso-path, a set of estimates for each value of $\gamma\in \Gamma=[0, \gamma_{max}]$. The maximal penalty $\gamma_{max}$ is the penalty level below which the first penalised group is let in to the model. In practice, we will use by default a 100 step log-linearly decreasing equidistant grid from $\gamma_{max}$ to $0.001\gamma_{max}$.  

Typically when using a lasso-based variable selection analysis one would use Cross Validation (CV) to choose the level of penalization that gives the best balance between model fit and prediction quality, or minimises the expected risk \citep[Sec. 5.2]{giraud2014}, also called the extra-sample error \citep[Sec. 7.1.]{Hastie2001}. In the point pattern context, \cite{Yue2014a} also chose their penalty using CV, but unfortunately no details were given how the data was partitioned or what error metric was used for the prediction. 
Conducting CV by splitting the constructed data frame that was used in the logistic regression part of the inference is not justified as the rows are dependent due to spatial correlation. This flaw leads us to over-estimate the complexity of the model.
Note also that it has been observed that as cross-validation is aimed at prediction, rather than model selection, and so too many variables are often retained~\citep[Section 2.5.1]{van2011statistics}. 

Let us define a CV procedure for spatial point pattern models that coincides with the CV for classical regression. Two elements are needed; a way to split the data into meaningful subsets of independent data; and a way to measure prediction error, or risk, with the already explained caveats. Let the observation window $W$ be partitioned into disjoint quadrats $W_1,...,W_K$, and write for each $k=1,...,K$
\[
\hat\theta_{\gamma, -k} = \text{argmax}_{\theta} L_\gamma(\x\setminus W_k;\theta)
\]
for the group lasso estimate using all the data {\em except} points inside $W_k$. For the CV risk we need to define prediction quality, for example the residual sum of squares used in linear regression. \cite{Baddeley2004} introduced the $h$-residual measure for point patterns which we can use to define a residual sum of squares. Define the CV $h$-residual as
\begin{eqnarray}
\label{eq:resid}
\hat R_{\gamma,k} &:=& R(W_k,\hat\theta_{\gamma,-k})\nonumber\\
&=&\sum_{u\in\x\cap W_k} h(u,\x\setminus u,\hat\theta_{\gamma,-k}) - \int_{W_k\times\{1,...,p\}} h(u, \x, \hat\theta_{\gamma,-k})\lambda_{\hat\theta_{\gamma,-k}}(u;\x) du\\
&=& \sum_{i=1}^p\left[\sum_{x\in\x_i\cap W_k} h((x,i),\x\setminus (x,i),\hat\theta_{\gamma,-k}) - \int_{W_k} h((x,i), \x, \hat\theta_{\gamma,-k})\lambda_{\hat\theta_{\gamma,-k}}((x,i);\x) dx \right],\nonumber
\end{eqnarray}
where $h$ is a non-negative function called the test function. \cite{Baddeley2004} and \cite{Coeurjolly2013b} list four options for the test function, of which we will look at three. The \emph{raw} residuals with $h=1$ assess only the trend part of the model, and the \emph{inverse} residuals with $h(u;\x,\theta) = \lambda_{\theta}(u;\x)^{-1}$ assess the interactions. The \emph{Pearson} residuals with $h(u;\x,\theta)=\lambda_{\theta}(u;\x)^{-1/2}$ is similar to the inverse but with variance that is in theory independent of $\lambda$ for the Poisson process.

We define the $K$-fold CV estimator of the prediction risk as the mean of the squared residuals
\[
\hat R_{CV}(\gamma) := \frac{1}{K}\sum_{k=1}^K \hat R_{\gamma,k}^2
\]
and we choose the minimizer of the estimated risk,
\[
\hat\gamma_{CV} := \text{argmin}_{\gamma\in \Gamma}\ \hat R_{CV}(\gamma),
\]
as the cross-validated penalty level.
Note that one can easily weight the residuals by relative importance of each type or the quadrat size. 

Figure \ref{fig:CVexample} in the appendix illustrates this approach for an example pattern generated by Experiment 1. The individual quadrats' residuals vary---this is not unexpected as the number of data points per quadrat is small---but the minimum average risk still leads to a reasonable penalisation.

In a CV procedure the model needs to be fitted $K$ times so 
computational cost and CV risk estimate stability are to be balanced. In the spatial setting an additional problem arises due to border correction, {\em viz.} we use $W_{k}\ominus b(o,r_{bor})$ instead of $W_{k}$ in (\ref{eq:resid}) with some $r_{bor}>0$. Since each sub-window needs to be reduced by the same border correction range as the original window to get truly independent subsets of data, the loss of data for estimating $\hat{R}_{\gamma,k}$'s limits the number of splits that can be done on $W$, as some data is simply lost in the process. For example, in a square window, a border correction range that is 5\% of the window's side together with a $3\times 3$ partitioning will effectively lead to a 50\% loss of data when estimating the risk  (cf. Appendix Figure \ref{fig:CVborder}). Furthermore, the varying abundances of the types need weighting to account for importances of different types, and heterogeneity of data should be considered as well. It is therefore very hard to give general guidelines for partitioning the window that would work in all scenarios. 



For comparison with the CV penalty selection, we also include a rule-of-thumb penalty selection based on the AIC. We computed the AIC for the group lasso as described by \cite{Breheny2009}. In our experiments, penalties with the lowest AIC, say $\gamma_{AIC}$, consistently led to too dense solutions. This suggests that either the pseudo-likelihood approximation or the spatial dependency, or both, lead to underestimation of the effective number of parameters. As a plug-in rule of thumb, initial trials indicate that a penalty around $(\gamma_{AIC}+\gamma_{max})/2$ leads to reasonable penalisation; we will report these results with the label AIC0.5. 

We also constructed a Bayesian algorithm for computing approximate posteriors $Pr(e_g=1|\x)$. We call this tailored algorithm VBSS (Variational Bayes Spike and Slab). It uses a quadratic approximation to the logistic function \citep{Jaakkola2000}, selects variables via the spike and slab priors \citep{Mitchell1988} and uses the mean-field variational approximation to infer the maximum-a-posteriors \citep{Ormerod2010}. In Experiment 1 we also checked a non-grouped version of VBSS and compared it to the MCMC based SSGAM \citep{Scheipl2011} that has a slightly different implementation of the spike and slab priors. All of these alternatives solve the same optimisation problem as group lasso (Equation \ref{eq:glasso}), but with different choices of penalisation.

\section{Alternative approach for multi-type interaction detection: Non-parametric Monte Carlo}
\label{sec:MC}
The current state-of-the-art methodology for analysing high dimensional multivariate point pattern data is based on non-parametric summary statistics and Monte Carlo (MC) testing. This approach splits the analysis into a set of bivariate tests for no co-associations (second order interactions). As a reference to our proposed method we apply a MC testing scheme in the BCI rain forest -like, large inhomogeneous simulation experiment of Section \ref{sec:exp5} and the actual BCI rain forest data analysis Section \ref{sec:bci}.  We give here a short description of MC testing but the interested reader should seek more in-depth texts on the topic see e.g. \cite{Illian2008a} Sec. 7.5; \cite{Baddeley2014,Velazquez2016, Brown2016}. 

Monte Carlo -based testing procedures in general, and tests of spatial type-to-type interactions in particular, are built on three components: 1) A null hypothesis that describes a counter factual pattern without the interactions of interest. 2) A statistical summary that measures the interaction of interest. 3) A statistical test based on the summary for measuring data's departure from the distribution of the summary when the null hypothesis holds.

In the context of potentially inhomogeneous patterns, a popular null model, especially in ecology, is some form of the inhomogeneous Poisson process (IPP)~\citep{Wiegand2012}. A common, asymmetric version can be simulated by keeping type $i$ locations fixed while sampling type $j$ locations from IPP independently from all other data. The idea is to nullify short range interactions but keep longer range, environment-related associations via the spatially-varying intensity. The intensity surface needed for the IPP simulations of 
$\x_j$ is estimated with a fixed smoothness that reflects understanding of the division between "short" and "long" ranges. We implement the IPP simulations following \cite{Wiegand2012}: For each type $i=1,...,p$, an intensity field $\eta_i(u), u\in W=[0,1000]\times[0,500]m$ is kernel-estimated on a $2\times2m$ grid using the border-corrected Epanechnikov kernel with bandwidth fixed to $30m$ (this is of course not optimal for all types; we simply emulate \citealt{Wiegand2012}). Then $n_i$ points are distributed on $W$ with density relative to $\eta_i(u)$, i.e. we use a conditional IPP as is commonly done. We simulate 999 patterns this way for each type $i=1,...,p$.

For the statistical summary of pattern interaction, a popular choice is the cross-Ripley's $K_{ij}$-function. It describes the number of type $j$ neighbours for an average point of type $i$, over different neighbourhood ranges. So, assuming isotropy, for each pair of bi-variate patterns the $K_{ij}(r)$-function estimate is a sampled curve over spatial scales $r>0$. In our analyses, we estimate the cross-type $K_{ij}(r)$ curves on a range grid $r=r_1,...,r_{max}$ with translation edge correction. For each species $i=1,...,p$ and $j=1,...,p$, we estimate the curve $K_{ij}^0=\{K_{ij}^0(r_1),...,K_{ij}^0(r_{max})\}$ from the bivariate data pattern $\x_i\cup \x_j$. When $i=j$ we estimate the univariate Ripley's $K_{ii}=K$. Subsequently, we estimate the curves $K_{ij}^b$ from the bivariate synthetic patterns $\x_i\cup \x_j^b$ where $\x_j^b$ is a simulation of the null model for $j$ as described above, and $b=1,...,999$. With the set of curves $(K_{ij}^0,...,K_{ij}^{999})$ for all $p^2$ combinations of $i$ and $j$, we do the variance stabilising $\sqrt{K(r)/\pi}$ transform to increase statistical power. 

The third component of the MC testing framework is then needed for a proper combined test for the $K_{ij}$ curves, i.e. to determine if $K^0_{ij}$ is different from the null model curves. A family of such tests are called deviation tests or envelope tests \citep{Myllymaki2016, Baddeley2014}, and several options are available. We will use the Studentised deviation test, which measures the $L^2$-distance of the Studentised curves (scaled with respect to the null model), and the rank envelope test, which is a multidimensional analogue of a rank-test ({\em cf.} \citealt{Myllymaki2016}). Each test leads to a $p$-value per test, say $p_{ij}$. We then report the values $e_{ij} := 1(p_{ij} < .05)$ as indicators of interaction.

This concludes our description of  our methodological framework, and explains our automated approach to the selection of  ``active'' interactions that are  important to explain the observed spatial pattern.

\section{Simulation trials}
To check the model fitting procedure before data-analysis we illustrate its characteristics via multiple simulation trials. We will focus on the estimation of the cross-interaction terms, which we can write as a square matrix
\[M = [M_{ij}]_1^p, \quad \text{where} \quad M_{ij} = \left\{\begin{array}{rl}
e_{g(ij)} & i\neq j\\
e_{g(i)} & i = j\\
\end{array}\right. .\]
We call $M$ the interaction matrix. Note that the coefficients related to each interaction function $\beta_{ij}^\T\g_{ij}$ are in groups, meaning that either the whole step-function is estimated to be $\hat\beta_{ij}=0$, or one or more of the components $\beta_{ijk}$ is estimated to be non-zero. This is an example of the group lasso~\citep{yuan2006model}.

The key quality metrics we will look at to assess group level performance are the true positive rate (TP) and the false positive rate (FP), stratified to intra- and inter-type interactions when relevant. 

We simulated patterns in either $W=[0,10]^2$ or $W=[0,1000]\times[0,500]m$ window with different settings for interactions. Experiments 1,2,4 study the method under the scenario of a correctly specified model, i.e. we simulate and estimate the multi-range multivariate model. In Experiments 3 and 5 we simulate Cox models to see how general and flexible the step-function (shrinkage) approach can be for interaction discovery. Simulations of our model were carried out using the Birth-and-Death algorithm for Figure \ref{fig:example1}, but for the trials we will use a Metropolis-Hastings algorithm with fixed point counts 
to keep intensities at desired levels \citep[for more detail we refer to][p. 147-154]{Illian2008a}.


\subsection{Experiment 1: Interactions in a small pattern}
We first simulated a small ($p=4$) example to familiarise the reader with our methodology and check that the method works as intended. We simulated 100 realisations of the multi-range multivariate saturation model with per type point counts $(100,100,50,150)$ in a $[0,10]^2$ window. The saturation levels were all set to $c=1$, producing low levels of interaction. Intra-type ranges were set to $\rv_i=(.1, .2, .3)$, and inter-type ranges were set to $\rv_l=(.1, .4)$. Types 1 and 2 were set to exhibit internally a mixture of short range repulsion and medium range clustering ($\beta_{ii}=(-1,1,0), i=1,2$), type 3 had some medium range clustering ($\beta_{33}=(0,1,0)$), and type 4 had no internal correlation. A positive inter-type correlation ($\beta_{ij}=(.6, .3)$) was set between types 1 and 2 and types 3 and 4. The true range vectors were used for fitting. Strongly penalising hyper-prior $Pr(e_g=1)\sim Beta(.1, 10)$ was chosen for the Bayesian methods. A $4\times 4$ partitioning of $W$ was used for CV to keep the data loss around 50\%.

\begin{figure}
\centering
\includegraphics[width=1\linewidth]{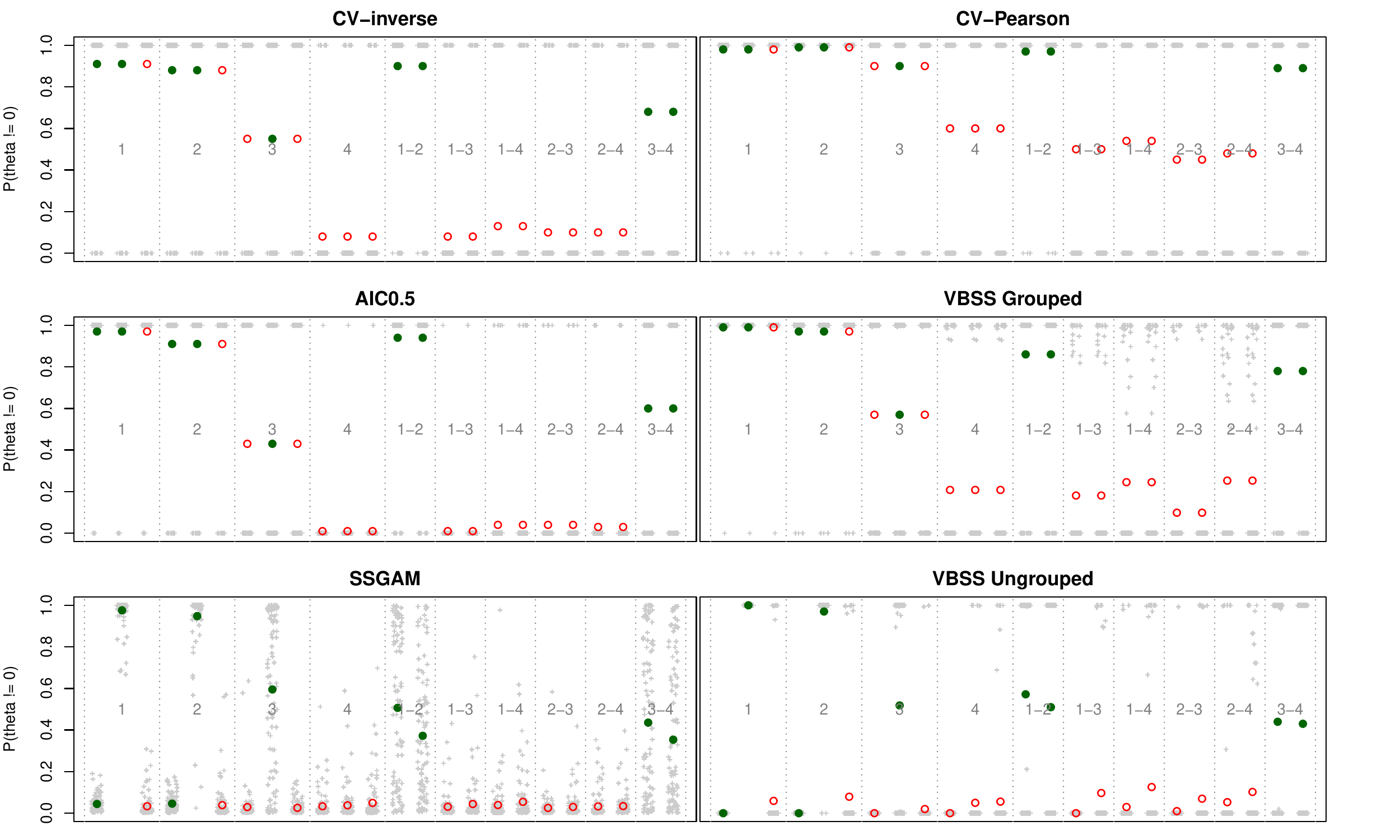}
\caption{Illustration of the variable selection results for experiment 1. Three different fitting algorithms (Group lasso, SSGAM and VBSS, as described in Section \ref{sec:penalised}). 
Penalty selection for  Group lasso based on CV with inverse (CV-inverse) and Pearson residuals (CV-Pearson) and AIC-based rule-of-thumb penalty (AIC0.5, see text). SSGAM and VBSS with $Beta(.1,10)$ penalty priors. ``Grouped'' and ``ungrouped'' refers to how the spike-variables (the non-zero variables) were defined in VBSS fits. Symbols denote coefficient-wise averages over 100 simulations, jittered grey crosses in the background denote the per simulation estimates. If a large symbol is closed, the true coefficient was non-zero. The grouping is denoted by vertical lines, the corresponding type(s) with numbers (e.g. intra-type interaction for type $i=3$ by "3", inter-type interaction for $i=1,j=3$ by "1-3" ) .}
\label{fig:exp1-methods}
\end{figure}

Figure \ref{fig:exp1-methods} depicts the average detection rates per coefficient $\beta_{ijk}$. Apart from Pearson residuals, the rates of the grouped methods are very similar, indicating the estimated effects do not depend on the chosen algorithm (group lasso or spike and slab). 
The non-zero structure in types 1 and 2 are detected well by the method. The un-grouped SSGAM and VBSS 
capture well the medium range clustering for types 1 and 2 but not the short range regularity. This indicates that the medium range clustering is the main signal at the group level. The type 3 sub-patterns had interaction only at medium range, and it seems to be difficult to uncover by any of the methods used. We posit that this difficulty is due to a smaller point count and mixed interaction with type 4. The grouping helps discover the inter-type interactions between the 1-2 and 3-4 pairs, evident from the lower detection rates of the un-grouped methods.

A summary of the true and false positive rates at group level for all methods are given in Table \ref{tab:experiment1}. The un-grouped outputs are considered non-zero per group if any of the group members were estimated non-zero. We used a .5 threshold for the Bayesian posterior probabilities for classification. The raw and Pearson residuals clearly are too prone to false positives in this example to be useful. 

\begin{table}
\caption{\label{tab:experiment1} Experiment 1 true positive and false positive rates, mean and (sd) over 100 simulations.}
\centering
\footnotesize
\setlength{\tabcolsep}{0.3em}

\begin{tabular}{r|ccccccc}
\hline
 & AIC0.5 & CV Inverse & CV Pearson & CV Raw & SSGAM & VBSS Grouped & VBSS Ungrouped \\ 
\hline
TP & .77 (.18) & .78 (.26) & .95 (.15) & .99 (.08) & .78 (.17) & .83 (.17) & .80 (.16) \\ 
FP & .03 (.07) & .10 (.20) & .51 (.32) & .94 (.16) & .02 (.06) & .21 (.19) & .12 (.15) \\ 
\hline

\end{tabular}
\end{table}

We also studied how varying the range vectors and dummy intensities affect the outcome. The maximum ranges in the simulations were $R=.3$ for intra- and $R=.4$ for inter-interactions. To simulate miss-specification of the ranges, 
we altered these by multiplying by a factor .5 or 1.5, and created range vectors for fitting with 1,2,3,4 or 5 equidistant steps. Then we fitted the model using the group lasso with the AIC0.5 rule for penalty with dummy intensity factor  $\rho$ either 2, 4 or 8. 

Two main features emerged from these studies (cf. Appendix Figure \ref{fig:exp1-ranges}). First, choosing ranges that are too short reduces the detection quality quite markedly. Second, the number of grid steps in the interaction functions can be misspecified without large variation in the results, giving a type of numerical robustness. Many small steps in the interactions functions 
is not recommended, due to the small number of pairs in the data that hit every annuli. For dummy intensity factor $2$ the joint point configuration did not have enough point pairs to fill every bin in the range grid of 5 steps, even after re-sampling the dummies (10 repeated attempts). As we also saw in the results for type $i=3$ above, using many steps may result in lower true positive rates at group level because the group lasso penalises the whole group equally over all its members. The false positive rates are higher for dummy intensity factor $\rho=8$ than for $\rho=4$, a side effect of the AIC0.5 rule-of-thumb. A check of the AIC curves showed that the minimum was often achieved with lower penalty when $\rho=8$, likely due to there being more observations in the logistic regression design matrix, an unsurprising indication that the standard model selection tools are not to be trusted when using pseudo-likelihood. Since we are focusing on the grouped analysis in this study we continue with group Lasso and VBSS.

\subsection{Experiment 2: Finding interactions in blocks}
Next, we increased the type count to $p=10$, and added interactions in two blocks with in-block pairwise interactions but no interaction between the blocks. The main task was to discover the two blocks with minimal amount of false positives in their cross-section. 

The ranges for simulating from the multi-range saturation model were all set to a two step vector $\rv=(.25, .50)$, and saturations fixed to $c=1$. The first block of 5 types had short and medium range clustering, $\beta_{ii}=(1,.5)$ for $i=1,...,5$, the second block of 5 types had short range repulsion with mild medium range clustering $\beta_{ii}=(-1,.5)$ for $i=6,...,10$. The inter-type interaction for each pair in both blocks was positive correlation with $\beta_{ij}=(.5, .25)$, with no correlation between blocks. We simulated three intensity scenarios, having point count per type $n_i$ either $50, 100$ or $200$, so that total point counts per simulated pattern were either $500$, $1000$ or $2000$.  The window was again $[0,10]^2$.

Fitting was done with the misspecified range vectors $\rv=(.15, 0.3)$, to increase the challenge (cf. Experiment 1). 
CV was conducted with a 5$\times$5 partitioning to keep expected data loss around 50\%. 
VBSS was fitted with three hyper-priors $\pi_l\sim Beta(.,.)$ ranging from flat (1,1) to medium (.1, 1) and strong (.1, 10) preference for no interactions. The choices correspond to increasing the penalisation in group lasso and facilitate comparisons between the methods.

The rates of interaction detection are shown in Table \ref{tab:experiment2}. VBSS produces many false positives, even with the strong prior that should penalise towards sparsity. 
The raw CV method works well, as does the inverse CV method, but the Pearson CV has a high false positive rate. 
AIC0.5 over-penalises, thereby missing all of the intra-type interactions, indicating that the rule-of-thumb is not generally useful. 

\begin{table}
\caption{Experiment 2 interaction detection rates, stratified by intra- and inter-type interactions and point-count per type. Mean over 50 simulations, with the standard deviation given in parenthesis. Intra-FP's are 0 by design. Lasso penalty selected with 25-fold CV.}
\label{tab:experiment2} 
\centering
\scriptsize
\setlength{\tabcolsep}{0.5em}
\begin{tabular}{r|lll|lll|lll}
\hline
 & \multicolumn{3}{c}{$n_i=50$}& \multicolumn{3}{c}{$n_i=100$}& \multicolumn{3}{c}{$n_i=200$}\\
Method & intra TP& inter TP & inter FP & intra TP & inter TP & inter FP & intra TP & inter TP & inter FP \\ 
  \hline
raw & .51 (.20) & .97 (.04) & .04 (.06) & .65 (.21) & .99 (.02) & .03 (.05) & .89 (.10) & 1.0 (.00) & .20 (.23) \\ 
  inverse & .41 (.22) & .96 (.05) & .04 (.07) & .63 (.37) & .95 (.11) & .15 (.27) & .17 (.33) & .76 (.19) & .06 (.21) \\ 
  Pearson & .90 (.13) & .99 (.02) & .55 (.36) & .97 (.05) & 1.0 (.01) & .78 (.35) & .99 (.04) & 1.0 (.00) & .90 (.27) \\ 
  AIC0.5 & .00 (.01) & .61 (.11) & .00 (.00) & .00 (.00) & .64 (.10) & .00 (.00) & .00 (.00) & .57 (.07) & .00 (.00) \\ 
  VBSS (1,1) & .85 (.12) & .95 (.04) & .50 (.13) & .97 (.07) & .99 (.02) & .91 (.20) & .91 (.09) & 1.0 (.00) & .86 (.25) \\ 
  VBSS (.1,1) & .37 (.16) & .58 (.27) & .23 (.12) & .97 (.08) & .85 (.19) & .87 (.22) & .90 (.11) & 1.0 (.02) & .46 (.48) \\ 
  VBSS (.1,10) & .09 (.11) & .14 (.22) & .07 (.07) & .98 (.07) & .62 (.19) & .77 (.24) & .84 (.20) & .97 (.07) & .34 (.45) \\ 
   \hline
\end{tabular}
\end{table}

\begin{figure}[!ht]
\centering
\includegraphics[width=.7\textwidth]{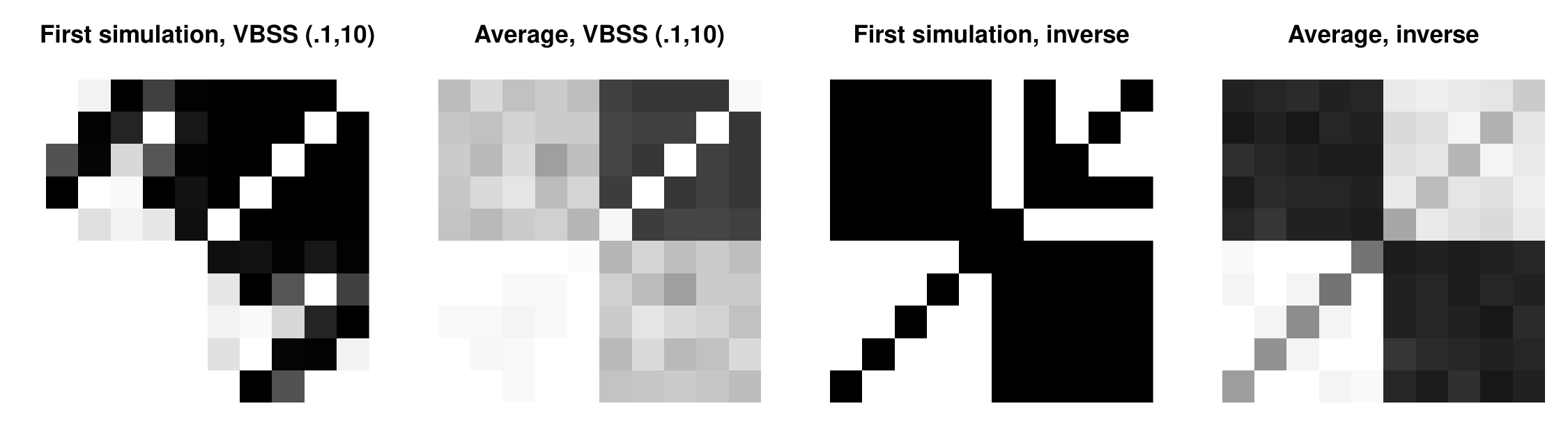}
\caption{Results for Experiment 2. Shown are an example interaction matrix estimate and the mean interaction matrix, for both VBSS and inverse CV methods. Point count per type was 100. Each pixel $(i,j)$ corresponds to $M_{i,j}$. Black means interaction detection rate was 0, and white means it was 1.}
\label{fig:experiment2}
\end{figure}

Figure \ref{fig:experiment2} depicts the interaction matrix estimate for a single realisation and the mean interaction over the simulations when $n_i=100$, as given by VBSS with the strong prior and inverse CV method. 
We plot the main diagonal of the interaction matrix from the southwest corner to the northeast corner, situating the $(1,1)$ pixel at the southwest corner, as per Figure 1 in~\cite{Flugge2014}. Note that white indicates that the detection rate was unity.

We note that the
first block is clear in VBSS matrices but the second block, with intra-type regularity mixing with the inter-type clustering, is not so clear and many false positives have been detected. The group lasso is not so efficient in detecting the intra-type interactions but the inter-type interactions are much better detected, with only a few false positives on average.

Further tailoring of ranges will improve quality, as per Experiment 1. Adjusting the priors helps VBSS but provides little practical gain over the automatic penalty rules for the group lasso which work reasonably well on all cases. As the priors lead to varying results, analysis with VBSS needs to include a sensitivity analysis, similar to the lasso penalisation selection. The major issue is that instead of having one tuning parameter there are several, and cross validating the space becomes infeasible as (at least with our implementation) the VBSS algorithm is not much faster for a single prior choice than the group lasso algorithm is for the whole penalty path. The usual Bayesian benefits such as posterior distributions and variances are not worth pursuing either due to the likelihood approximation. For these practical reasons we will continue with only the group lasso to perform computational feasible variable selection.

\subsection{Experiment 3: Detecting blocks of interacting Log-Gaussian Cox processes}
Next we check if interactions can be detected in correlated log-Gaussian Cox process data. We do not expect a large power here due to  the local Poisson distribution of the patterns, but it is important to check how flexible the detection is under model misspecification as the rainforest data potentially exhibits a variety of spatial mechanisms.

\begin{figure}[ht!]
\centering
\includegraphics[width=1\linewidth]{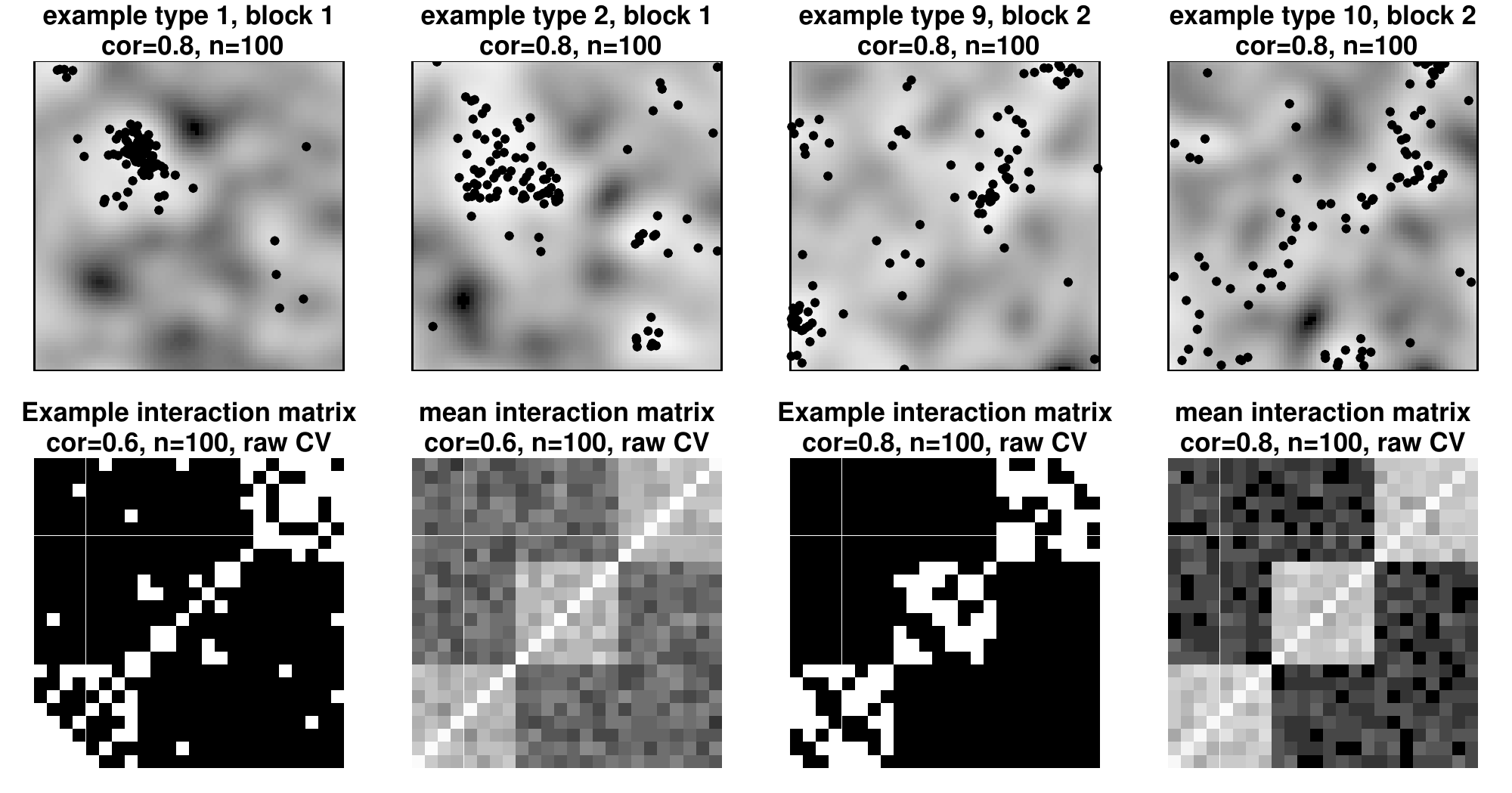}
\caption{Experiment 3, block correlated log-Gaussian Cox processes. Top row: Examples of the point patterns in a simulation, two patterns from each of the two intra-correlated blocks, overlaid on their generating log-Gaussian fields. Bottom row: Example and the average interaction matrix for two levels of intra-block correlation, estimated by the raw residual CV method; black means interaction detection rate was 0, and white means it was 1. Note that the $(1,1)$ pixel is located at the southwest corner of the image.}
\label{fig:exp4-example}
\end{figure}

We simulated homogeneous, stationary multivariate log-Gaussian Cox processes of $p=24$ types, structured into 3 correlated blocks of 8 types each. Inside a block the 8 latent Gaussian fields are correlated LMC fields \citep{Gelfand2002}, for which we set the cross-field correlation levels to either $.6$ or $.8$. All fields were driven by a Mat\'{e}rn covariance function with smoothness $\nu=10$, marginal variance $\sigma^2=2$ and correlation range $3$ (so that $cor(r>3)<.1$). For illustration, four sub-patterns from two different blocks are plotted on the top row of Figure \ref{fig:exp4-example}, overlaid on their generating fields. We ran the experiment with constant $n_i=50$ and $n_i=100$ points per type. The window was $[0,10]^2$.

\begin{table}
\caption{\label{tab:exp4}Experiment 3, log-Gaussian Cox processes, inter-type interactions, average and sd over 50 simulations. }
\centering
\setlength{\tabcolsep}{0.5em}
\hspace*{-.5cm}\begin{tabular}{r|rr|rr|rr|rr}
\hline
  &\multicolumn{4}{c}{corr $.6$}   & \multicolumn{4}{|c}{corr $.8$}\\
    &\multicolumn{2}{c}{$n_i$=50}     &\multicolumn{2}{c}{$n_i$=100}     &\multicolumn{2}{|c}{$n_i$=50}    &\multicolumn{2}{c}{$n_i$=100}\\
Method & TP & FP  & TP & FP  & TP & FP  & TP & FP \\ 

  \hline
raw & .60 (.12) & .16 (.16) & .60 (.16) & .16 (.20) & .62 (.13) & .08 (.14) & .65 (.12) & .07 (.15) \\ 
  inverse & .60 (.10) & .14 (.12) & .64 (.10) & .17 (.11) & .67 (.08) & .11 (.13) & .73 (.13) & .17 (.15) \\ 
  Pearson & .75 (.11) & .41 (.24) & .71 (.09) & .31 (.18) & .77 (.08) & .32 (.19) & .80 (.07) & .31 (.20) \\ 
  AIC0.5 & .20 (.06) & .00 (.00) & .22 (.06) & .00 (.00) & .23 (.07) & .00 (.00) & .27 (.08) & .00 (.00) \\ 
   \hline

  
   \end{tabular}
\end{table}

Estimation range vectors for intra- and inter-type interactions were set to $\rv=(.25, .5)$ after examining one realisation's cross pair correlation function. The CV's were carried out using a $3\times 3$ partitioning to keep the expected data loss at 50\%. 

Table \ref{tab:exp4} lists the detection rates. The block structure is detected to some extent, with somewhat elevated false positive rates. The raw residual CV performs the best, Pearson CV resulting in high false positive rates, and inverse CV landing overall somewhere in between. The AIC0.5 rule penalises too much to detect more than approximately 25\% of the block structure. As expected, an increase in inter-field correlation improves the detection, as does doubling the point count. The bottom row of Figure \ref{fig:exp4-example} depicts the estimated interaction matrices for the raw CV when point counts were $n_i=100$. Even though the inter-type interactions are not always detected correctly, the intra-type clustering are detected quite well. 

We noticed that improvements are easily achieved by adjusting the range vectors used in the fitting; but eye-balling a single realisation's pair correlation function is obviously not optimal for a repeated experiment. A practical solution is to choose the ranges per realisation based on a proper exploration of the data, in which case we expect quite a good performance considering how misspecified the model is for the generating mechanism. During our development of the methods, further 
improvements in this experiment were also achieved by increasing the level of clustering in the patterns, either by increasing the variance or by decreasing the range. For example, setting $\sigma^2=3$ (the original level was $\sigma^2=2$) reduces false positive rates and increased true positive rates up to 10\% across the table. The presented results were chosen as a generic illustration, and, by eye at least, the simulation parameters produced clustering similar to that observed in the rainforest data.

\subsection{Experiment 4: Rare interactions}
To focus more specifically on the false positive rates of the candidate methods, we simulated patterns where, at most, there is only one pairwise inter-type interaction that is non-zero. We considered four scenarios: (1) homogeneous multivariate Poisson patterns; (2) homogeneous multivariate Poisson patterns with an extra intra-correlated pattern; (3) inhomogeneous multivariate Poisson patterns, and (4) inhomogeneous multivariate Poisson patterns with an extra intra-correlated pattern. 

For this and the following Experiment 5 we used real covariate maps from the BCI dataset. Thirteen different soil variables (such as soil Ph and magnesium concentration), together with elevation and elevation gradient maps, are available on a $20\times20m$ grid covering the observation window $W=[0,1000]\times[0,500]m$. For this experiment (and the data analysis later on) we decomposed the 15 covariate maps using a singular value decomposition of the point-wise measurement matrix, and kept the six component maps corresponding to the six largest singular values, capturing approximately 70\% of the features. These six pointwise independent PCA maps were then used as covariates. 

To generate trends for the inhomogeneous multivariate Poisson processes, we combined the PCA covariates linearly with random coefficients taking values $-1$, 0 or $+1$, with probabilities $.25, .5,$ and $.25$, respectively. 
For each simulation of the multivariate process, two trend maps were generated and exponentiated to work as intensity surfaces. Each of the two intensity surfaces was assigned to half of the $p$ sub-processes. This way each simulated multivariate pattern has two blocks of types, $p/2$ each, with very strong intra-block correlation that is fully explained by the covariates. For each simulation the point counts per type were set to range from $50$ to $300$, increasing log-linearly so that lower counts were more common. We repeated the experiment for $p=10,20$.

The extra intra-correlated patterns were generated with the multi-type multi-scale model with homogeneous intensity of 100 points, and exhibited very short range repulsion and medium range clustering ($\rv=(1, 20), \beta=(-10, 1), c=(3, 3)$). For fitting we set all ranges to $\rv=(7,15)m$.

The results for the four scenarios are given in Table \ref{tab:experiment3}. With the Poisson data, the methods all produce some false positives. The inverse CV is best, with around 6\% false positive rates. With the exception of Pearson CV, including the extra interacting type sharpens the results by reducing the false positive rates. Every method discovers the extra interaction perfectly. Including covariate dependency, and producing correlation, does not increase the false positive rates overall. Again, the inclusion of a single interacting type reduces the false positive rates while itself being clearly detected. Doubling the type count roughly quadruples the parameter count, but it does not seem to affect the quality as false positive rates go down, albeit not always by a factor of 4.

\begin{table}
\caption{\label{tab:experiment3}Experiment 4, detection rates with and without habitat effects and mostly no interaction. When type count $p$ is doubled, the inter-pair count is quadraupled ($\times$ 4.22)	.}
\centering
\setlength{\tabcolsep}{0.5em}

\begin{tabular}{c|cccc|cccc}
\hline
& \multicolumn{4}{c}{$p=10$} & \multicolumn{4}{c}{$p=20$} \\
\hline
 & raw & inverse & Pearson & AIC0.5  & raw & inverse & Pearson & AIC0.5 \\ 
\hline
&\multicolumn{8}{c}{Homogeneous}\\
\hline
FP intra & .44 (.42) & .07 (.12) & .08 (.15) & .15 (.12) & .19 (.20) & .02 (.04) & .04 (.07) & .08 (.06) \\ 
FP inter & .47 (.39) & .06 (.09) & .12 (.19) & .18 (.07) & .22 (.19) & .03 (.04) & .07 (.10) & .10 (.03) \\ 
\hline
&\multicolumn{8}{c}{Homogeneous + 1 interaction}\\
\hline
FP intra & .13 (.27) & .01 (.03) & .14 (.17) & .00 (.00) & .08 (.20) & .04 (.10) & .06 (.10) & .00 (.00) \\  
FP inter & .07 (.13) & .00 (.01) & .13 (.16) & .00 (.00) & .06 (.17) & .02 (.06) & .06 (.07) & .00 (.00) \\ 
TP & 1.0 (.00) & 1.0 (.00) & 1.0 (.00) & 1.0 (.00) & 1.0 (.00) & 1.0 (.00) & 1.0 (.00) & 1.0 (.00) \\  

\hline
&\multicolumn{8}{c}{Inhomogeneous}\\
\hline
FP intra & .25 (.36) & .11 (.13) & .19 (.34) & .14 (.14) & .13 (.09) & .00 (.02) & .02 (.03) & .10 (.08) \\ 
FP inter & .22 (.34) & .08 (.10) & .18 (.31) & .13 (.10) & .10 (.07) & .01 (.01) & .01 (.01) & .06 (.03) \\ 
\hline
&\multicolumn{8}{c}{Inhomogeneous + 1 interaction}\\
\hline
FP intra & .08 (.19) & .02 (.06) & .13 (.22) & .00 (.00) & .12 (.24) & .00 (.00) & .04 (.05) & .00 (.00) \\ 
FP inter & .05 (.08) & .01 (.02) & .13 (.20) & .00 (.00) & .12 (.21) & .00 (.01) & .04 (.06) & .00 (.00) \\ 
TP & 1.0 (.00) & 1.0 (.00) & 1.0 (.00) & 1.0 (.00) & 1.0 (.00) & 1.0 (.00) & 1.0 (.00) & 1.0 (.00) \\ 

\hline
\end{tabular}

\end{table}

In the estimation procedure we did not penalise the covariate coefficients, because we found that penalising them using the group lasso lead to under-prediction of the trend due to the shrinkage effects of lasso, consequently leading to under-penalisation by the cross validation. When the covariate effects are to be considered more closely we suggest a bias correction step to the lasso or the use of a less strongly penalizing added term such as the SCAD or MCP \citep{Breheny2009}.

\subsection{Experiment 5: Independent patterns in a rainforest landscape}
\label{sec:exp5}
As a final synthetic example we simulated $p=64=4\times 16$ independent, inhomogeneous patterns in $W=[0,1000]\times [0,500]m$ window with first order interaction depending on the covariates in the BCI data. We selected half ($2\times 16$) of the patterns to come from a Thomas cluster process \citep[][Sec. 6.3.2]{Illian2008a} with two different dispersal ranges and points-per-cluster rates, to arrive at an appropriate degree of heterogeneity. We set the first block's model to generate patterns with a few large clusters ("Thomas 1"), and the second block's model to generate patterns with many small clusters ("Thomas 2"). The second half of the patterns was generated using the multi-type multi-range model with either repulsion followed by clustering ("Geyer 1") or just repulsion ("Geyer 2"). The ranges of the multi-range models were inversely dependent on target point count: This way small amount of points could spread out more realistically, and if target point count was high then maximal packing density would not be violated (i.e. only a finite amount of "discs" with certain radius fit inside $W$). 

To simulate habitat effects we connected the patterns to the maps of four covariates $Mn, P, pH$ and $grad$ which have been found to be relevant covariates for the rainforest population \citep{Schreeg2010}. The covariate values were first standardized. Then for each simulated species we sampled uniformly $t\in\{0,..,4\}$ of them, and summed them point-wise with weights $(t,...,1)/(t+1)$ to produce a surface. The surface was then used as the unscaled intensity field for Thomas processes generator points, and as first order field for the multi-step multivariate Geyer models. The covariate choices and the coefficients were kept fixed over the replicates of the multivariate simulations.

In a single realisation of the process, each of the 4 models were simulated 16 times, with target intensities within each 16 type block ranging from 50 to 1000 points, median 225. The realised point counts ranged from 34 to 1024 due to edge effects while simulating the Thomas process (points outside $W$ were dropped). The point count in a single multivariate realisation was around 20800. 

\begin{figure}[!h]
\includegraphics[width=1\textwidth]{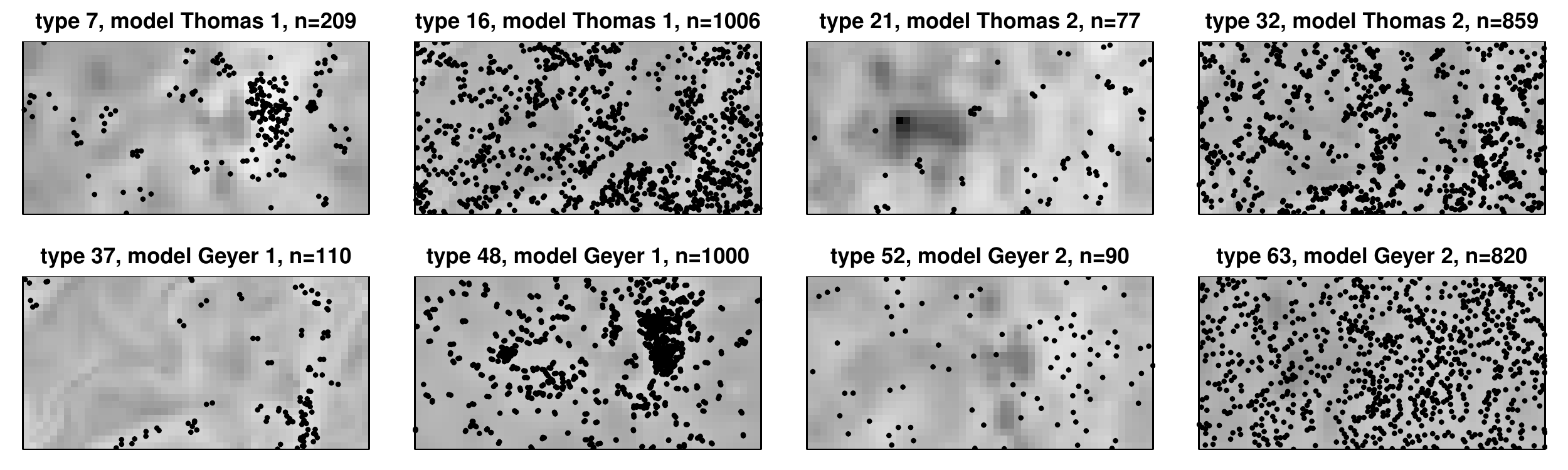}
\caption{Experiment 5, examples of the sub-patterns with non-zero habitat association in a single multivariate realisation, together with the habitat effect field. Window $[0,1000]\times [0,500]m$.}
\label{fig:experiment5-patterns}
\end{figure}

Figure \ref{fig:experiment5-patterns} depicts 8 sub-patterns in a typical realisation, each overlaid on its inhomogeneity generating habitat field. "Thomas 1" patterns have about half the number of clusters of the "Thomas 2" patterns. 
Note that it is hard to see from the plots the very short scale features operating at ranges 1-10m as the area is so large. 

For fitting we used ranges $\rv=(10,20)$ and 7$\times$4 partitioning to keep the estimated data loss at 50\%. Table \ref{tab:experiment5} shows the detection rates for the proposed methods together with the MC based Studentised deviance and rank envelope test, both based on the cross-$K$ functions estimated over ranges .5-15m and 1-50m (corresponding to short and medium-long range testing scenarios) with 999 simulations of the IPP null-model, as described in Section \ref{sec:MC}. 
Apart from the Pearson CV, and the AIC0.5 rule-of-thumb which penalises too much, the detection rates are good. Around or below 5\% of the false inter-type interactions are detected while 74-96\% of the intra-type interactions are detected. The model based and the Monte Carlo based estimates are very close to each other. 
 

\begin{table}
\caption{\label{tab:experiment5}Experiment 5, synthetic forest of independent species. 
Detection rates for the new approach, together with results for two Monte Carlo tests, Studentised deviance ("St'd dev.") and rank envelope ("Rank"), with two different testing ranges. By design, all 64 intra-type interactions were non-zero, and all 2016 inter-type interactions were zero. Average and sd over 10 simulations.}
\setlength{\tabcolsep}{0.5em}
\centering
\begin{tabular}{rr|rrrr}
\cline{1-6}
& & CV-raw & CV-inverse & CV-Pearson & AIC0.5 \\ 
  \cline{2-6}
Gibbs model&  TP & .76 (.34) & .91 (.06) & .96 (.08) & .35 (.06) \\ 
     &  FP & .01 (.01) & .05 (.10) & .20 (.12) & .00 (.00) \\ 
   \hline
& & St'd dev. (.5-15m) & St'd dev. (1-50m) & Rank (.5-15m) & Rank (1-50m) \\ 
  \cline{2-6}
Monte Carlo& TP & .74 (.05) & .89 (.03) & .80 (.05) & .96 (.01) \\ 
 &FP & .02 (.00) & .00 (.00) & .03 (.00) & .02 (.00) \\ 
   \hline

\end{tabular}

\end{table}

\begin{figure}[!h]
\includegraphics[width=1\textwidth]{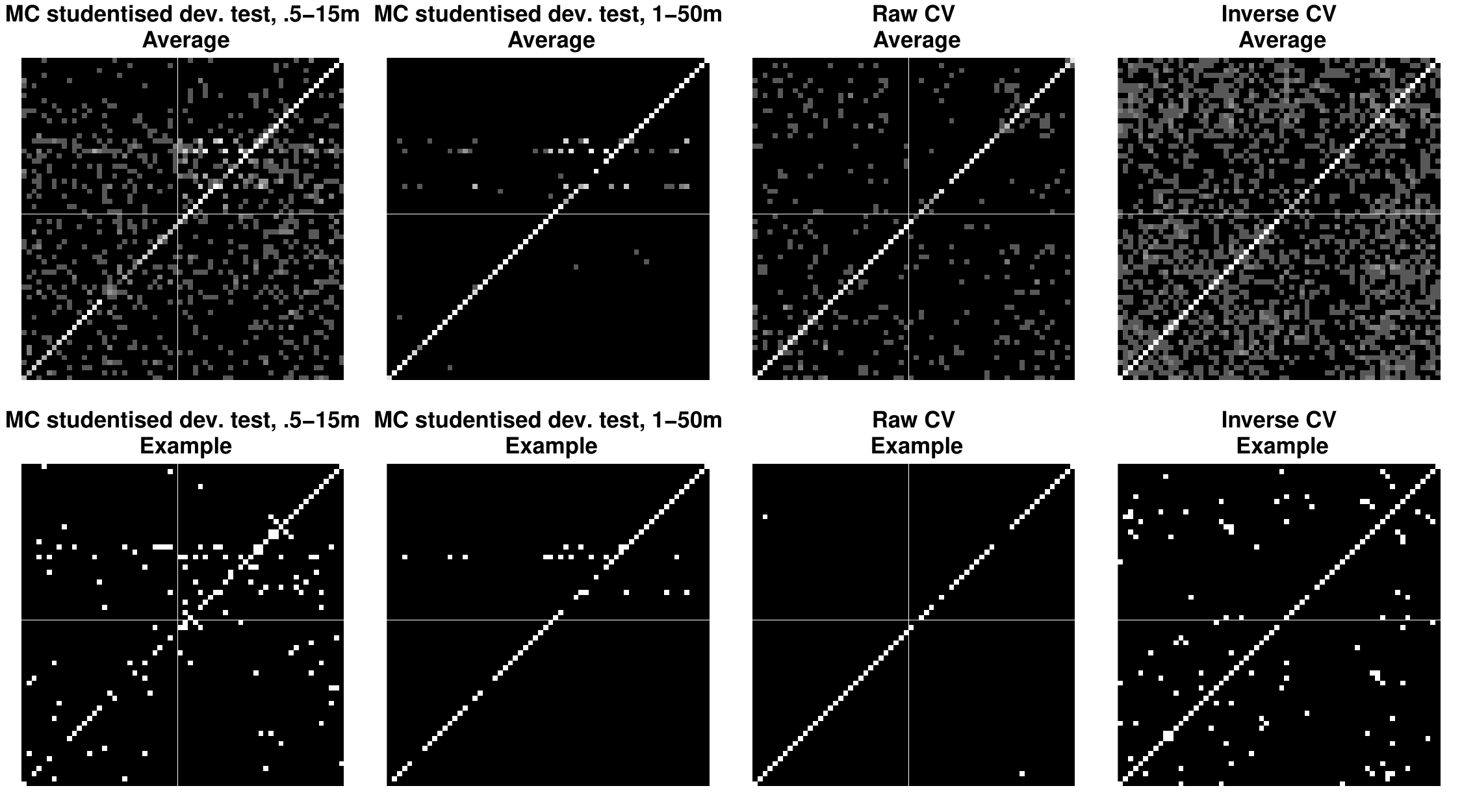}
\caption{Experiment 5, averages (top row) and examples (bottom row) of estimated interaction matrices. The results from two Monte Carlo tests with two different range-scale coverages (see text) and the raw and inverse CV methods are shown. Perfect estimate would be a diagonal matrix. Thomas models are in the lower left section and Geyer models in the top right section of each matrix.}
\label{fig:experiment5}
\end{figure}

Somewhat surprisingly, the Thomas patterns, for which the model is miss-specified, are not the harder of the two families to discover. 
Figure \ref{fig:experiment5} shows the averages and examples of interaction matrix estimates for the Monte Carlo tests and raw and inverse CV methods. We placed the $(1,1)$ interaction at the southwest corner of each image. The matrices are ordered by model (Thomas 1, Thomas 2, Geyer 1, Geyer 2) and increasing point count from bottom left to top right within the model blocks. For intra-type interactions (on the diagonal), the rarer species' interaction within the Geyer 2 model seem to be harder to discover with the CV methods. This could be due to the fixed $\rv$ that was used for estimation when the processes had varying ranges. The MC null-hypothesis design, where the types corresponding to columns in the interaction matrix were kept fixed and the types corresponding to rows were randomised, leads to non-symmetric estimates. This is an important feature of the MC method, and as we return to in the data analysis below, can lead to interpretation issues. The Geyer 1 block, where there is short scale repulsion, and medium scale attraction, is the major source of false positives for the MC tests, and types with higher abundance show generally higher false positive rates. This is potentially important, because such a mixture of repulsion and attraction is likely to be common in many plant communities.

\subsection{Summary of the simulation trials}
The results from the simulation studies are very encouraging despite the sometimes large dimensionality of the underlying problems, especially demonstrated with Experiment 5. In Experiment 1 we established feasibility of our model, and the agreement between penalisation methods validate the approach. Experiment 2 demonstrated the model's ability to discover block community structures in a moderately complicated 10 type pattern. Experiment 3 showed that the model can detect significant interactions in not just realisations of itself, but also in realisations of Cox processes. Experiment 4 showed that false positive rate does not need to be high for us to be able to discover rare events. Finally in Experiment 5 we saw that the method handles rainforest-like patterns with covariate effects, varying point counts and various interaction mechanisms and models well. 

The experiments suggest that CV with Pearson residuals is not reliable in practice. The raw residuals work well in some cases (Experiments 2 and 3), but fails completely in others. The inverse residuals produced most consistently good detection. When the raw and inverse residuals both worked well, the variance of the inverse residuals were higher, so computing both is advisable. 

As noted in the Experiments, we did not specifically tailor the range parameters for each fitting task per se. Adjusting the model's range-scales according to specific analysis is highly recommended as they are expected to improve the results in practice.

\section{Data example: Rainforest interactions}
\label{sec:bci}
The BCI rainforest dataset consists of multivariate point patterns corresponding to censuses of rainforest plants (shrubs and trees) living in a $W=[0,1000]\times[0,500]m$ area of Barro Colorado Island, Panama. Censuses have been taken regularly since 1981 \citep{Hubbell2005, Condit1998, Condit1999}. In each census, woody plants (shrubs, trees) with diameter at breast height (dbh) over 1cm were cataloged, noting their dbh, location, species, condition and some other details not relevant to our studies in this article. 

The total number of species is about 300, with slight variation over the years due to immigration and extinctions in $W$, which is physically an open area allowing for migration . For the example data analysis we chose the 2005 census, and selected some specific species. Recent studies \citep{Kanagaraj2011, Yang2016a} suggest that spatial features of the plants vary with life stage. Normally the distribution of young plants
is more clustered due to seed dispersal and adult plants' distribution more regular due to competition based self thinning, and that resource requirements also change with maturation. We therefore selected a subset of species for which an estimate of reproducible size, a surrogate for juvenile/adult life stage, was available \citep[unpublished data by R. Foster, available as supplements for][]{Flugge2014}. We included only adult plants and excluded species with less than 50 adults in the region, leading to a multivariate pattern of $p=83$ species. 
Point counts vary from 50 to 8784, with a median of 118 and a total of 31650. 

\subsection{Interaction detection using the introduced methods}
We carried out the interaction estimation with a multi-range multivariate saturation model using group lasso penalisation as in Experiment 5. Habitat effects were accounted for by including the 6 PCA maps as covariates as described in Experiment 4 (using PCA maps avoids problems with co-linearity of covariates). We set the range vector equal across all 3486 intra- and inter-type interactions, and fitted the model twice, with range vector $\rv = (7, 15)m$ and with range vector $\rv=(7, 15, 30)m$, the ranges being concordant with previous results of neighbourhood dependent growth models within the BCI forest (Table 4, \citealt{Uriarte2004}). 
We used the dummy intensity factor 4, with a minimum of 500 dummies per type to avoid singularities. The saturation parameters are set according to Section \ref{seq:2.2}. 
To correct for edge effects we implemented a 30m buffer zone, and, to keep expected data loss below 50\%, the CV partitioning was fixed to 6$\times$3. For comparison, we also implemented the MC tests as described in Section \ref{sec:MC}.

\subsection{Results}
\begin{figure}
\centering
\includegraphics[width=1\linewidth]{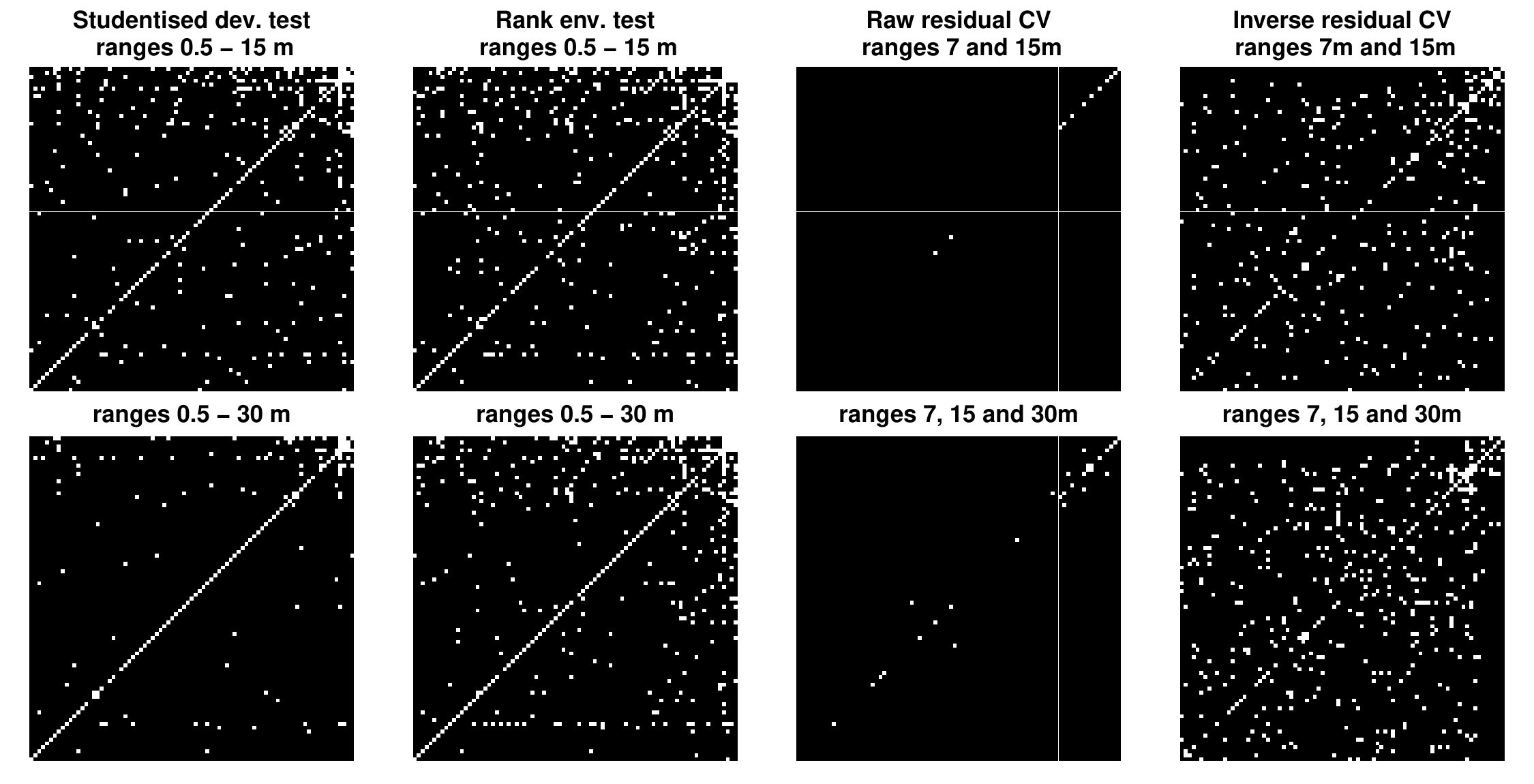}
\caption{Different estimates of the interaction matrix of the adult plant BCI '05 census. {Columns 1 and 2:} Non-parametric Monte Carlo method with two different MC tests and two different spatial range vectors. The species kept fixed in the test is on the x-axis, the randomized species on the y-axis. \emph{Columns 3 and 4:} Group lasso with raw residual CV and inverse residual CV methods, and two range vectors. Species are arranged by point count, increasing from left/bottom to right/top.}
\label{fig:bci-matrices}
\end{figure}

The resulting interaction matrices for the two MC tests are depicted on the first two columns of Figure \ref{fig:bci-matrices}. The group lasso estimates with raw and inverse residual CV penalty selection are depicted on the two rightmost columns. The top row results involve two ranges up to 15m, while the bottom row involves three ranges up to 30m. The MC test interaction matrices are not symmetric by design. For example, according to the rank envelope test at ranges .5-30m, species $i=10$ interacts significantly with 1 species when the other species in a pair is randomized in the test, but it interacts significantly with 22 species when itself is the one randomised in the test. Although, biological interactions (competition) can be asymmetric, with one species being a superior competitor to a second species, the spatial correlations that emerge from these interactions should be symmetric. The non-symmetric spatial interaction matrices produced by the MC method are therefore problematic when it comes to interpretation. In comparison, the inverse CV model-based method estimates the number of interactions for species $i=10$ to be 5, of which 4 are amongst the 22 species indicated by the MC test. Another difference is that the MC test outcome suggests that more abundant species interact more. The model based results do not indicate such a trend (cf. Appendix Figure \ref{fig:bci-abundance}); so when the number of inter-type interactions are of similar order, the distributions are different. Finally, nearly all species are deemed internally structured by MC, but around or less than half by the model-based approach (Table \ref{tab:bci-rates}).  

\begin{table}
\caption{\label{tab:bci-rates}Intra- and inter-type interaction rates estimated with the proposed method using raw and inverse CV, and Monte Carlo tests using the Studentised deviance ("St'd dev.") and rank envelopes ("rank"). Results for range-scales upto 15m and 30m separately. $p=83$ species, 3403 distinct pairs.}
\centering
\setlength{\tabcolsep}{0.5em}

 \begin{tabular}{lrrrr}
\hline
  & St'd dev. & Rank & CV-raw & CV-inverse \\ 
  \hline
  Up to 15m & & & &\\
    \hline
intra & 0.89 & 0.88 & 0.14 & 0.55 \\ 
  inter & 0.05 & 0.06 & 0.00 & 0.04 \\ 
   \hline
  Up to 30m & & & &\\
  \hline
intra & 0.95 & 0.96 & 0.33 & 0.60 \\ 
  inter & 0.02 & 0.05 & 0.00 & 0.06 \\ 
   \hline
\end{tabular}
\end{table}

We have designed a complete analysis pipeline that can be used to answer a number of specific ecological questions, and in particular uncover important processes affecting the forest. For example, we could now proceed by analysing the covariate effects (Appendix Figure \ref{fig:bci-covariates}) to see how important environment (soil, elevation) effects are in structuring the forest. To then consider structure caused by second order effects, we could further study the interaction matrices to discover sub-communities and interacting groups as was done by \cite{Flugge2014} and \cite{Morueta-Holme2016} based on different MC approaches.  

\section{Discussion}
We have developed a flexible model-based method for detecting small scale interactions in multivariate spatial point patterns, demonstrated its potential on several synthetic examples, and applied it to a large rainforest data set. 
This is in contrast to using log-Gaussian Cox processes for multivariate modeling, as in for example~\cite{Waagepetersen2015b}, where small scale interactions are harder to capture.
In so-doing we have greatly extended the potential applicability of model-based inference for multitype point patterns, allowing both for the analysis of more interactions, and more complex multiscale behaviour.

In the Thomas cluster process, and in Cox processes in general, the two-stage generation of "offspring" given "mothers" is a natural model for plant regeneration in natural plant populations. However, as the offspring are assumed mutually independent, the model is suitable for only a particular type of pattern, such as just germinated individuals. The individual-to-individual competition and survival over a plant's lifetime prior adulthood is better captured with explicit point-wise interaction models, such as the Gibbs models. Alternative models that emulate natural thinning in addition to natural clustering can be constructed \citep{Stoyan1979, Andersen2015, Lavancier2016}, but fitting such models rely heavily on non-parametric methods for which simultaneous multivariate analysis is currently not well understood. Our approach takes a more data-analytical approach by including all effects simultaneously and addressing the high dimensionality with penalised inference. 

The model can be tailored for specific applications with relatively low effort since apart from the point-to-point indicator functions the analysis pipeline is fixed. For example, adding Strauss components might detect clustering more efficiently as the saturation level of the Geyer components caps the clustering tendency. Such a model would not work as a generative model and simulations would not be stable (singular clusters would form, \citealt{Gates1986}), but it could be effective as a statistical interaction discovery tool. 

The pipeline is also computationally practical. Group lasso algorithms are very fast and work well with computational underlying calculations for sparse matrices, which the approximation design matrices for Gibbs models are by construction. On a regular laptop it took us days to compute all MC estimates to compare to our model, but it took only hours to estimate the full model. Of course, one can parallelise the MC estimation since the type-to-type interaction tests are performed independently.

An advantage of our model-based approach is the explicit treatment of covariates that may affect the intensity but not the interactions directly. This means it is possible to infer which covariates are important for the distributions of the different species. In contrast the MC approach that predominates the ecological literature relies upon inhomogeneous Poisson processes to capture the first order (environmental) processes, and the user needs to determine/estimate the scale over which these processes are affecting the intensities. The current trend has been to use one scale for all species, but this is unlikely to be optimal. However, we note that such covariates are not always available, and the MC method might be preferable when covariates have not been measured but are thought to be important. The within-type spatial structure that is commonly found in the data is also broken down when conducting the null model test in the MC approach, and it is not clear what effect this might have on results. Nonetheless the user also has to make important decisions in our model-based approach, such as the range of scales over which to carry out the analysis, and as we discuss below the model validation and penalisation methods also are an area of future exploration.

The computational burden of model fitting was increased tenfold because we determined the best penalty level using cross validation. To simplify the penalty selection, we tried including a few extra Poisson "noise" sub-patterns into the data to select the penalisation so that the noise stayed independent. The result of this approach was inconclusive, and remains an area of future investigation. 
We also tried using cross validation with the constructed design matrix of the logistic regression likelihood, as is common for the the lasso applications, but much like with the AIC, too many false positives resulted for this approach. Standard model selection approaches do not do well because the likelihood is an approximation, rather than the correct form of the likelihood. Correcting the score and Hessian of the pseudo-likelihood while doing penalisation would be useful in this regard, and some work has already been done in the unpenalised case by \cite{Coeurjolly2013}. Using a corrected Hessian would also allow us to do better inference on the covariates, as (approximate) confidence intervals could then be constructed.

Replacing the group lasso with some other penalisation could have several benefits. Some penalisations shrink non-zero coefficients less which is ideal both for the predictions needed in the cross validation step, and downstream analysis of the estimated effects. For example, we suspect the shrinkage issue prevented the method from working properly when covariates were penalised and consequently the inhomogeneous trends became too flat. A further limitation of the group lasso is that all group members were penalised equally, this in turn leading to low group detection with many range-steps and low amount of data per range annuli. A more refined group penalisation method 
such as sparse-group lasso or minimax concave penalty is more sensitive to individual group members being non-zero \citep{Breheny2009} and would allow for more detailed interaction functions with more steps. Adding steps does however require using more dummy points 
so that each annuli registers something and numerical problems are avoided. We could also replace the step functions with overlapping components, such as radial basis functions, and gain not only numerical stability but also smoother interaction function estimates. Further refinement to the CV optimisation penalty grid is also needed in actual applications. In our high-$p$ examples the 100 step grid was often too coarse to include/exclude individual groups.

The model is constructed via small point-to-point interactions, and any large scale unobserved variability is not equally well captured. It might be possible to add spatial random effects, such as Gaussian processes, to the first-order interaction and still use the same pipeline. This would bridge the interface between Gibbs models and log-Gaussian Cox processes, an exciting area of future investigation.  This paper therefore stands as a further step to understand  general heterogeneous and high dimensional point process observations.

We would also like to point out that the modular construction of the point-to-point interaction functions provides a potentially useful connection to dynamic modeling of ecological communities 
defined through generations of dispersal related birth and competition based death events \citep{Law2000, Law2003}. If the transition probabilities are modelled log-linearly, the stationary distribution (if it exists) of the corresponding birth-and-death process following the detailed balance condition would be a Gibbs process of the kind we have discussed here in the more general, potentially inhomogeneous and highly multivariate setting. 

\section{Acknowledgements}
The BCI forest dynamics research project was founded by S.P. Hubbell and R.B. Foster and is now managed by R. Condit, S. Lao, and R. Perez under the Center for Tropical Forest Science and the Smithsonian Tropical Research in Panama. Numerous organizations have provided funding, principally the U.S. National Science Foundation, and hundreds of field workers have contributed. We thank the EPSRC for support via EP/N007336/1 and EP/L001519/1, and S. C. Olhede also acknowledges support from the 7th European Community Framework Programme (a European Research Council Fellowship via Grant CoG 2015- 682172NETS).

We thank the two anonymous reviewers for their very helpful comments.

\section{Software and data}
All computations were implemented in R (v.3.3.1). The random fields for the log-Gaussian Cox models were simulated using the R-package "RandomFields" (v.3.1.3), and uniform simulation and some utilities were used from the R-package "spatstat" (v1.46.1). For the group lasso we adapted the local coordinate descent algorithm in the R-package "grpreg" (v.3.0-2) with the inclusion of the offset terms. The VBSS was implemented by hand. For the SSGAM we used the R-package "spikeSlabGAM" (v.1.1-11), with hyper-prior Beta-distribution settings as for the VBSS and otherwise using default parameters. An R-package implementing the method pipeline is available from the first author. 

The BCI data is available for research purposes from \\\texttt{http://ctfs.si.edu/webatlas/datasets/bci/}. 

\bibliographystyle{rss}
\bibliography{multigeyer}

\newpage

\appendix

\section{Details about the likelihood approximation}
\label{appendix:likelihood}
Let $X$ be the point process generating the pattern $\x$ and let $D$ be a marked dummy point process in the window $W$ which is independent of $X$ and has known intensity functions $\rho_i, i=1,...p$. \cite{Baddeley2014} proposed to solve likelihood optimization problem by using the estimating function 
\begin{equation}
s_W(X,D;\theta) := \sum_{u\in X\cap W}\frac{\rho(u)\vv(u; X\setminus u)}{\lambda_\theta(u; X\setminus u)+\rho(u)} - \sum_{u\in D\cap W}\frac{\vv(u;X)\lambda_\theta(u;X)}{\lambda_\theta(u;X)+\rho(u)},
\end{equation}
where $\rho(u) = \rho_i(x)$ for $u=(x,i)$. With the help of the Campbell formula and the Georgii-Nguyen-Zessin theorem \citep[formulas 1.5.10 and 6.6.2][]{Illian2008a}, it can be shown that $s_W$ is an unbiased estimation function and that finding the root of $s_W$ gives an unbiased estimate of the maximum of the likelihood (\ref{eq:log-linear}). 

The $s_W$ as a function of $\theta$ is the derivative of 
\begin{equation}
\log \tilde{f}_\theta(X\cap W) = \sum_{u\in X\cap W}\log \frac{\lambda_\theta(u; X\setminus u)}{\lambda_\theta(u; X\setminus u)+\rho(u)} + \sum_{u\in D\cap W}\log \frac{\rho(u)}{\lambda_\theta(u;X)+\rho(u)},
\end{equation}
which formally is the likelihood of a logistic regression with variables $\tau(u)=1(u\in X)$ for $u\in X\cup D$ and
\[
P(\tau(u)=1) = \frac{\lambda_\theta(u; X\setminus u)}{\lambda_\theta(u; X\setminus u)+\rho(u)} = \frac{\exp[\theta^\T\vv(u;X) + \log\rho(u)^{-1}]}{1+\exp[\theta^\T\vv(u;X) + \log\rho(u)^{-1}]} .
\]
In practice the method works as follows: We sample a set of dummy points $\psi=\{(x,t)\}=\cup \mathbf{\psi}_i$ with sub-patterns $\mathbf{\psi}_i$ having a known distribution with constant intensity $\rho_i$ in $W$, for each type $i=1,...,p$. 
We then calculate the vectors $\mathbf{b}(u):=\vv(u;\x)$ for each $u\in \x\cup\psi$. Let $N = \#(\x\cup \psi)$. The log-likelihood of the logistic regression can then be written in a compact vector form corresponding to
\begin{eqnarray}
\label{eq:approxlikelihood}
\log \tilde{f}_\theta(\x) &=& \sum_{u=(x,i)\in (\x\cup\psi)\cap W}\left\{t(u) (\theta^\T \mathbf{b}(u) + o(u) ) - \log[1 + \exp(\theta^\T \mathbf{b}(u) + o(u)] \right\}\\
&=& \mathbf{t}^\T(B\theta + \ov) - 1^\T_N\log[1_N + \exp(B\theta +\ov)],\nonumber
\end{eqnarray}
where $t(u) = 1(u\in \x)$ indicate the data points, $B$ is a row matrix of $\mathbf{b}(u)$'s for each $u\in\x\cup\psi$, and $o(u) = -\log \rho_{i}(u)$ are offset terms. 

\section{Derivation of saturation approximation of the Papangelou terms under independence}
\label{appendix:saturation}
We show how the probability of saturation of the components in the log-Papangelou conditional intensity can be derived using independence assumption. First we note that for a homogeneous Poisson process neighbour counts in any set depend only the volume of the set. Thus to simplify from the annuli notation it suffices to consider only $r=r_k > r_{k-1}=0$. Then
\begin{align}
\omega(u,\x) &= min[c, ne(u,\x)] + \sum_{x\in \x}\{min[c,ne(x,\x\cup u)]-min[c,ne(x,\x\setminus u)]\} \\
&= min(c, \#[b(u,r)\cap \x]) + \sum_{x\in\x}1[x\in b(u,r)]1\{\#[b(x,r)\cap \x]\in [0,c)\}
\end{align}
where $1()$ denotes the indicator function. Assume now that $\x$ comes from a Poisson point process with intensity $\lambda$. Write $a=\lambda|b(o,r)|$. The Georgii-Nguyen-Zessin formula \citep[][p.399]{Illian2008a} gives
\begin{align*}
\E\omega(u,X)&= \E\min(c,\#[b(u,r)\cap X]) + \E \sum_{x\in X}1[x\in b(u,r)]1\{\#[b(x,r)\cap \x]\in [0,c)\}\\
&= \E\min(c,\#[b(o,r)\cap X]) + a \E 1\{\#[b(o,r)\cap X]\in [0,c)\}
\end{align*}
Now, the random variable $y:=\#[b(o,r)\cap X]$ is $Poisson(a)$ distributed. Denote its CDF with $F_a$. Then
\begin{align*}
&= \sum_{k=0}^{c-1}P(y=k)k + c\sum_{k=c}^{\infty}P(y=k) \ \ + a\sum_{k=0}^{c-1}P(y=k)\\
&= a\sum_{l=0}^{c-2}P(y=l) + c[1-F_a(c-1)] + aF_a(c-1)\\
&= c[1-F_a(c-1)] + a[F_a(c-1)+F_a(c-2)],
\end{align*}
giving the function $t(c)$ in the text. Note that we also use the fact $\lambda \approx n/|W|$.

\section{Illustration of the interaction and potential functions}
\label{appendix:interaction}

\begin{figure}
	\centering
	\makebox{\includegraphics[width=1\textwidth]{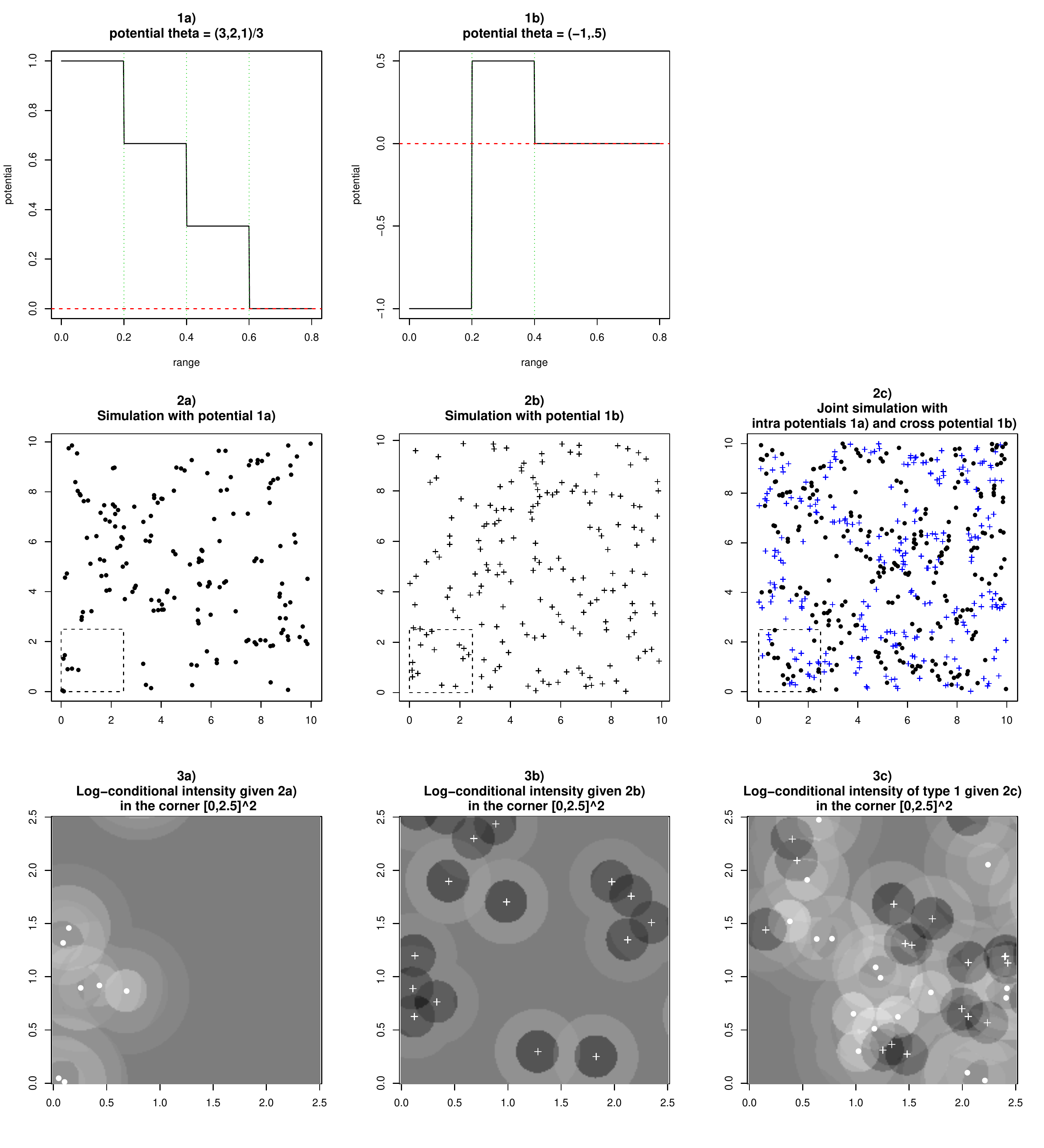}}
	\caption{\textit{Top row:} Two different potentials. Vertical lines mark the range vector stops. \textit{Middle row:} Simulations from the saturation model with $c=1$ and the two potentials, (a-b), and a joint bivariate simulation with intra-potentials given by 1a) and inter-potential given by 1b). \textit{Bottom row:} The conditional intensities of the saturation model given the simulations; in 3c) just for type $i=1$. Conditional intensities in log-scale. Background grey colour means 0 or no potential, darker colour means negative and lighter colour means positive potential.}
	\label{fig:example1}
\end{figure}

Figure \ref{fig:example1} (top row) depicts two particular potential shapes with different $\rv$'s and $\beta$'s, to illustrate their role. The first function corresponds to a decreasing attraction in range, and the second function has both a repulsion and an attraction component. How the attraction and repulsion impact the point pattern depends on the choice of the individual $g_{ijk}$'s: In this example  we use the saturation model $\g_{ij}$'s with $c_{ijk}\equiv 1$. The first two subfigures in the middle row of Figure \ref{fig:example1} show univariate simulations from each of the two interaction functions. The third subplot in the middle row shows a bivariate simulation with intra-type interaction given by the first potential and inter-type interaction given by the second potential. The bottom row shows the conditional intensity (\ref{eq:papangelou2}) at the window locations $u\in W$ given the simulated patterns.

In the case of the bivariate pattern the conditional intensity is for points of the first type. The conditional intensities exhibit various features: For the first potential function, high potential locations are near data-points, but only if the data-points do not already have neighbours (those data-points' neighbourhoods are already saturated). With the second potential any location too near the data-points has a low potential (repulsion), but being too far from the data-points is not encouraged by the potential either (attraction). In subfigure 3c) we see the complex mixture of first type's internal potential and the intra-type potential: A point of the first type would be welcome near a point of it's own kind (attraction) but unwelcome near a point of the second kind (repulsion).

\section{Additional figures}

\begin{figure}
\centering
\includegraphics[width=1\textwidth]{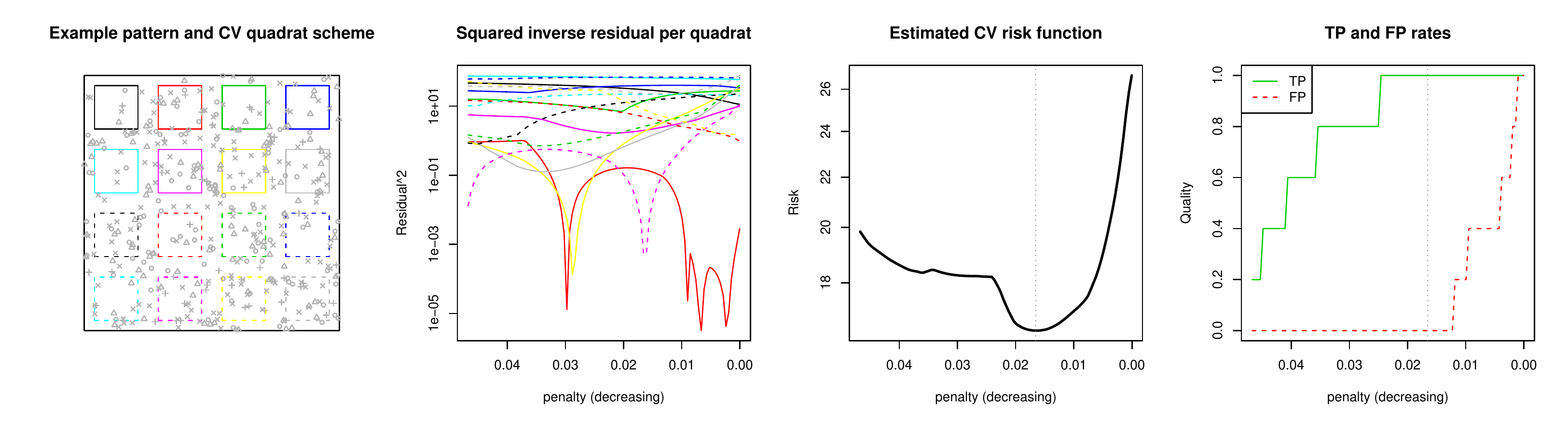}
\caption{Example of the CV penalty selection. From left: A pattern and the CV partitioning; The squared inverse residuals $\hat R_{\gamma,k}^2$ per quadrat $k=1,...,16$; The estimated prediction risk $\hat{R}_{CV}(\gamma)$; true and false positive rates per $\gamma$. Vertical line indicates the penalty level that would be chosen by the method.}
\label{fig:CVexample}
\end{figure}

\begin{figure}
\centering
\includegraphics[width=0.6\textwidth]{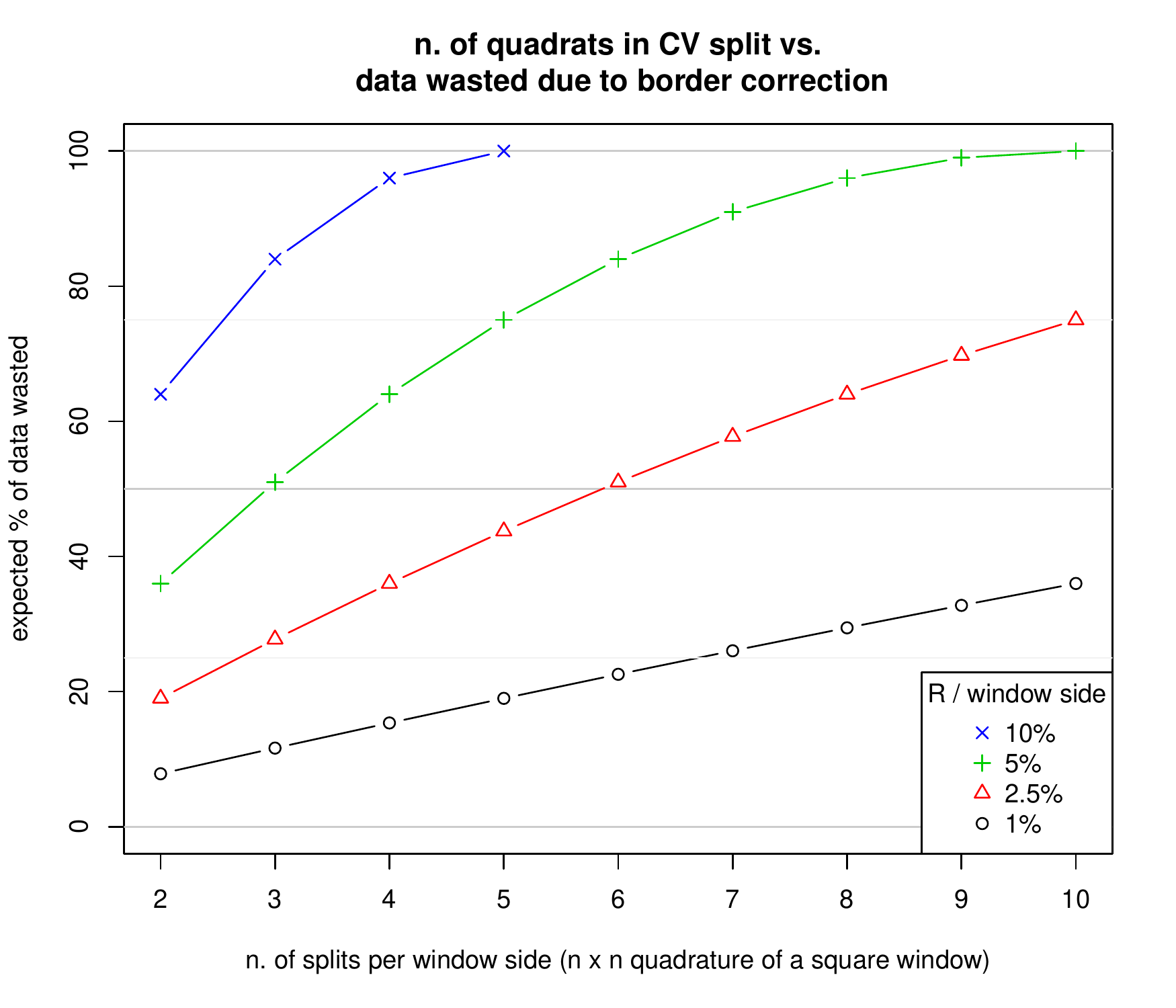}
\caption{Loss of data when doing CV splits with border correction with range $R$ in a square window.}
\label{fig:CVborder}
\end{figure}

\begin{figure}
\centering
\includegraphics[width=1\linewidth]{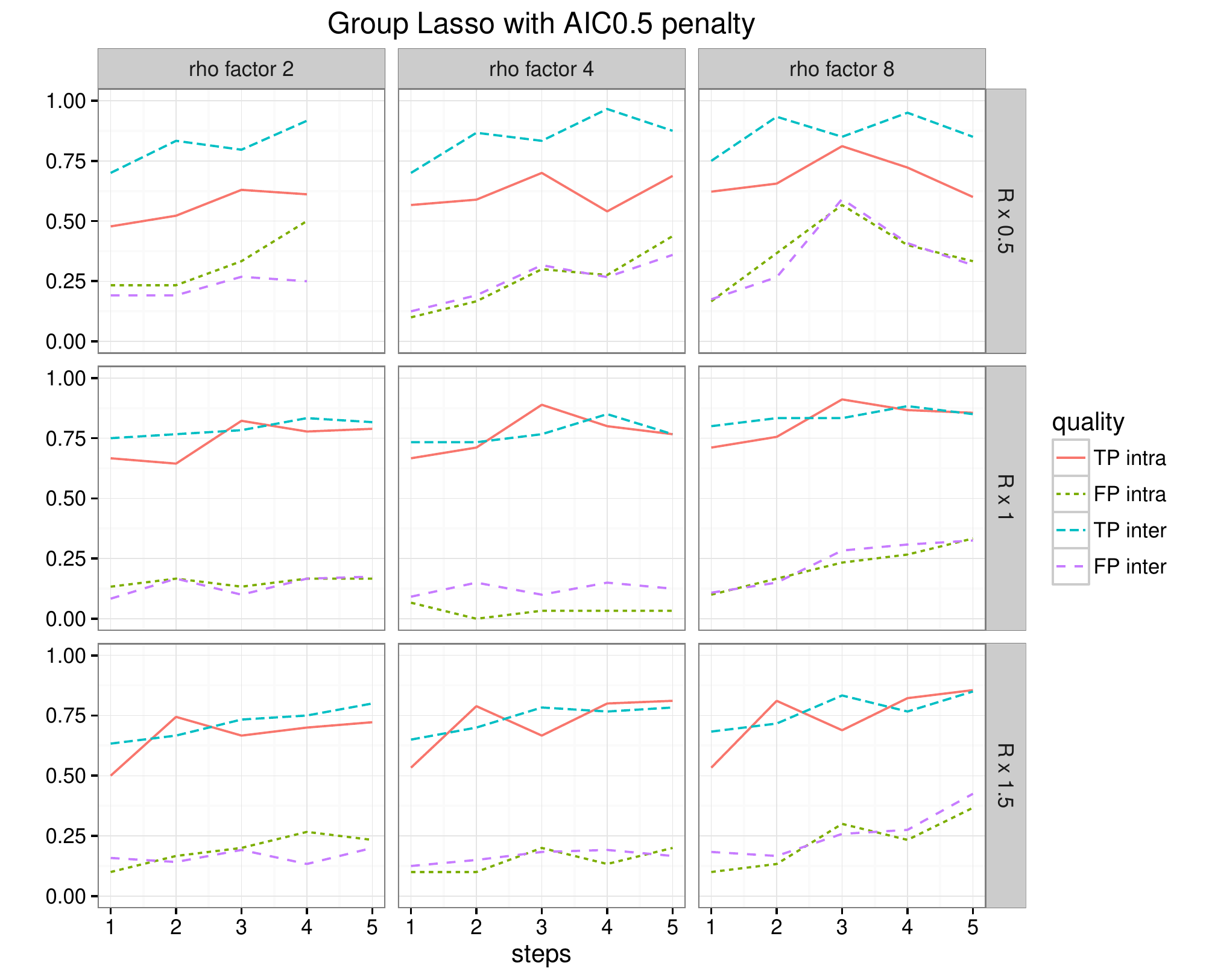}
\caption{Experiment 1, how does varying the range vectors $\rv=(r_1,...,r_{steps}=R_{max})$ and dummy intensity factor affect the quality of group lasso interaction detection quality. In the simulations the intra-type interaction maximum range was $R=.3$ and the inter-type $R=.4$, and the number of steps were 3 (intra) and 2 (inter). On the first row the estimation $R_{max}=.5 R$ and bottom row $R_{max}=1.5R$.}
\label{fig:exp1-ranges}
\end{figure}

\begin{figure}
\centering
\includegraphics[width=1\linewidth]{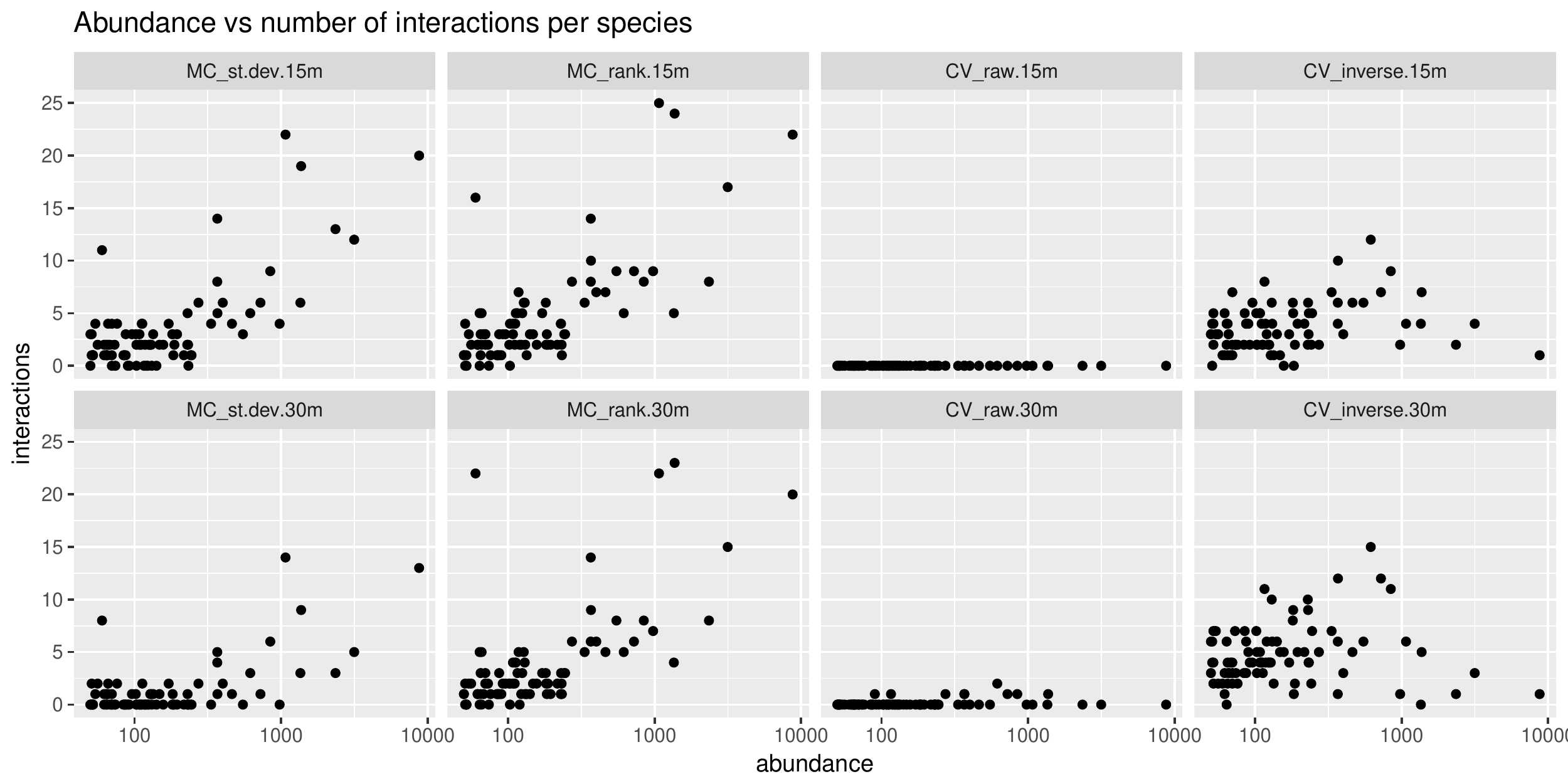}
\caption{BCI '05 adults analysis, number of detected inter-species $(ij)$ interactions for each species $i$, as a function of abundance. Two MC tests, two model based estimates; ranges up to 15m (top row) and up to 30m (bottom row).}
\label{fig:bci-abundance}
\end{figure}

\begin{figure}
\centering
\includegraphics[width=1\linewidth]{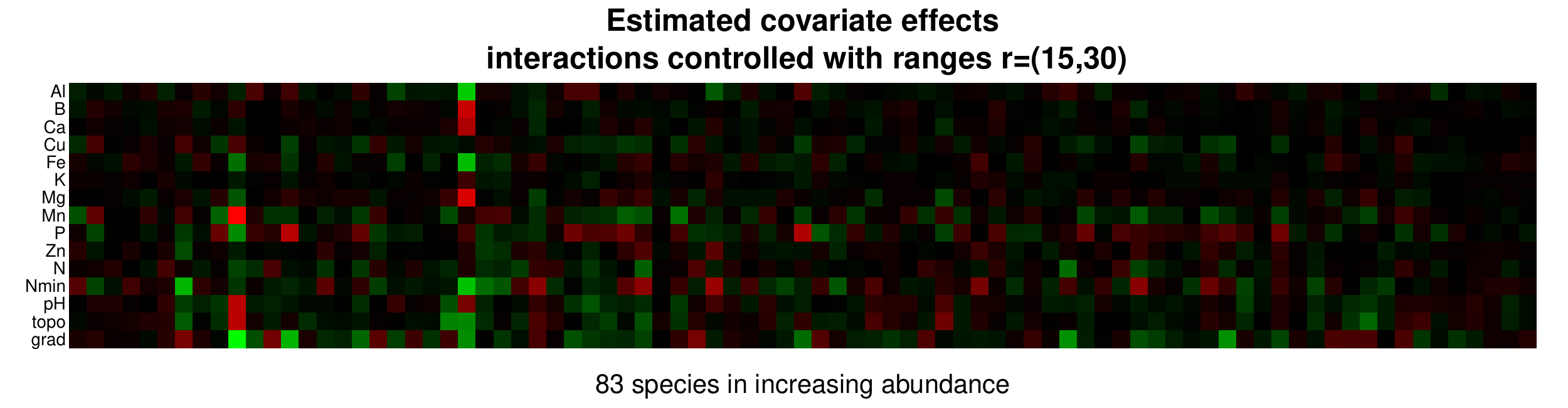}
\caption{The estimated covariate effects with the raw CV method when range vectors are $\rv=(15,30)m$. Positive connections green, negative in red, brightness reflects magnitude (black=0). Range of values [-0.9,1.03]. The estimates do not vary much between the methods or the used range vectors because the covariates were not penalised.}
\label{fig:bci-covariates}
\end{figure}


\end{document}